\documentclass{article}%
\usepackage{amsmath}
\usepackage{graphicx}
\usepackage{amsfonts}
\usepackage{amssymb}%
\newtheorem{theorem}{Theorem}

\newtheorem{corollary}[theorem]{Corollary}

\newtheorem{lemma}[theorem]{Lemma}

\newtheorem{proposition}[theorem]{Proposition}

\newenvironment{proof}[1][Proof]{\textbf{#1.} }{\ \rule{0.5em}{0.5em}}
\begin{document}

\title{The Complexity of Agreement}
\author{Scott Aaronson\thanks{University of California, Berkeley. \ Email:
aaronson@cs.berkeley.edu. \ Supported by an NSF Graduate Fellowship.}}
\date{}
\maketitle

\begin{abstract}
A celebrated 1976 theorem of Aumann asserts that honest, rational Bayesian
agents with common priors will never \textquotedblleft agree to
disagree\textquotedblright: if their opinions about any topic are common
knowledge, then those opinions must be equal. \ Economists have written
numerous papers examining the assumptions behind this theorem. \ But two key
questions went unaddressed: first, can the agents reach agreement after a
conversation of reasonable length? \ Second,\ can the computations needed for
that conversation be performed efficiently? \ This paper answers both
questions in the affirmative, thereby strengthening Aumann's original conclusion.

We first show that, for two agents with a common prior to agree within
$\varepsilon$\ about the expectation of a $\left[  0,1\right]  $\ variable
with high probability\ over their prior, it suffices for them to exchange
order $1/\varepsilon^{2}$\ bits. \ This bound is completely independent of the
number of bits $n$ of relevant knowledge that the agents have. \ We then
extend the bound to three or more agents; and we give an example where the
economists' \textquotedblleft standard protocol\textquotedblright\ (which
consists of repeatedly announcing one's current expectation) nearly saturates
the bound, while a new \textquotedblleft attenuated protocol\textquotedblright%
\ does better. \ Finally, we give a protocol that would cause two Bayesians to
agree within $\varepsilon$\ after exchanging order\ $1/\varepsilon^{2}%
$\ messages, and that can be \textit{simulated} by agents with limited
computational resources. \ By this we mean that, after examining the agents'
knowledge and a transcript of their conversation, no one would be able to
distinguish the agents from perfect Bayesians. \ The time used by the
simulation procedure is exponential in $1/\varepsilon^{6}$\ but not in $n$.

\end{abstract}

\section{Introduction\label{INTRO}}

A vast body of work in AI, economics, philosophy, and other fields seeks to
model human beings as \textit{Bayesian agents}---agents that start out with
some prior probability distribution over possible states of the world, then
update the distribution as they gather new information \cite{rn}. \ Because of
its simplicity, the \textquotedblleft
humans-as-roughly-Bayesians\textquotedblright\ thesis has remained popular,
despite the work of Allais \cite{ah}, Tversky and Kahneman \cite{tk}, and
others; and despite well-known problems such as old evidence \cite{earman}.
\ But one aspect of human experience seems especially hard to reconcile with
the thesis.

Pick any two people, and there will be some topic they disagree about:
capitalism versus socialism, the Israeli-Palestinian conflict, the
interpretation of quantum mechanics, etc.\footnote{If you disagree with this
assertion, you are simply providing further evidence for it!} \ The more
intelligent the people, the easier it will be to find such a topic. \ If they
discuss the topic, chances are excellent that they will not reach agreement,
but will instead become more confirmed in their previous beliefs. \ This is so
even if the people respect each other's intelligence and honesty.

The above facts are known to everyone, yet as Aumann \cite{aumann}\ observed
in 1976, they constitute a serious challenge to Bayesian accounts of human
reasoning. \ For suppose Alice and Bob are Bayesians, who have the same prior
probabilities for all states of the world, but who have since gained different
knowledge and thus have different posterior probabilities. \ Suppose further
that, conditioned on everything she knows, Alice assigns a posterior
probability $p$ to (say) extraterrestrial life existing. \ Bob likewise
assigns a posterior probability $q$. \ Then provided both agents know $p$ and
$q$ (and know that they know them, etc.), Aumann showed that $p$ and $q$ must
be equal. \ This is true even if neither agent has any idea on what sort of
evidence the other's estimate is based. \ For the sort of evidence can itself
be considered a random variable, which is ultimately governed by a prior
probability distribution that is the same for both agents.

Admittedly, the agents are unlikely to agree \textit{immediately} after
exchanging $p$ and $q$. \ For conditioned on Alice's estimate being $p$, Bob
will revise his estimate $q$, and similarly Alice will revise $p$ conditioned
on Bob's estimate $q$. \ The agents will then have to exchange their
\textit{new} estimates $p^{\prime}$\ and $q^{\prime}$, and so on iteratively.
\ But provided the set of possible states is finite, it is easy to show that
this iterative process must terminate eventually, with both agents having the
same estimate \cite{gp}. \ In conclusion, then, there is no reason for the
agents ever to disagree about anything!

On hearing this theorem for the first time, all of us come up with plausible
ways in which actual human beings might evade its conditions. \ People have
self-serving biases; they often discard or distort evidence that goes against
what they want to believe \cite{gilovich}. \ (According to an often-cited
study \cite{cross}, 94\% of professors consider themselves better than their
average colleagues.\footnote{This is logically possible, but one assumes the
response would be similar were the professors asked about their
\textit{median} colleagues.}) \ People might interpret the same assertion
differently. \ Or the assertion might be inherently ambiguous, if it deals
with beauty or morality for instance. \ People might weigh the same evidence
by different criteria. \ They might not understand the evidence. \ They might
defend their opinions as high-school debaters do, out of sport rather than a
desire for truth. \ They might not report their opinions with candor; or if
they do, they might not trust others to do likewise. \ 

In our view, the real challenge is not to list such caveats, but to sift
through them and to discover which ones are fundamental. \ As an illustration,
several of the caveats listed above disappear once we assume that all people
have a common prior. \ For among other things, such a prior would assign
common probabilities to all possible ways of parsing an ambiguous sentence,
and to all possible ways of weighing evidence. \ Understandably, then, much of
the criticism of Aumann's theorem has focused on the common prior assumption
(see \cite{aumann2,ch,gul} for a discussion of that assumption).

But suppose we accept that two people have different priors. \ The obvious
question is, \textit{what caused their priors to differ?} \ Different career
choices? \ Different friends? \ Different kindergarten teachers? \ Whatever is
named as the first influence, we need merely go back in time to before that
influence took effect. \ At the earlier time, the two people had the same
prior by assumption. \ So at later times, they would not really have different
priors, just different posteriors obtained by starting from the same prior and
then conditioning on different life experiences. \ If we push this reasoning
to its limit, as Cowen and Hanson \cite{ch} do, we are left wondering whether
prior differences could be encoded in DNA at conception. \ Even then, how much
confidence should you place in an opinion, if you know that were your genes
different, you would have the opposite opinion? \ More generally, on what
grounds can you favor your own prior over another's? \ For all you know, your
prior was \textquotedblleft switched by accident\textquotedblright\ with
someone else's at birth!

After staring into the metaphysical abyss of prior differences, the natural
reaction of a computer scientist is to step back, and ask if there is some
simpler explanation for why Aumann's theorem fails to describe the real world.
\ Recall that in the theorem, Alice's and Bob's opinions only became equal by
the end of a hypothetical conversation. \ Might that conversation last an
absurdly long time? \ After all, if Alice and Bob exchanged everything they
knew, then clearly they would agree about everything! \ But presumably they
are not Siamese twins, and do not have their entire lives to talk to each
other. \ Thus \textit{communication complexity} might provide a fundamental
reason for why even honest, rational people could agree to disagree. \ Indeed,
this was our conjecture when we began studying the topic.

\textit{Computational complexity} provides a second promising reason. \ If a
\textquotedblleft state of the world\textquotedblright\ consists of $n$ bits,
then Aumann's theorem requires Alice and Bob to represent a prior probability
distribution over $2^{n}$\ possible states. \ Even worse, it requires them to
calculate expectations over that distribution, and update it conditioned on
new information. \ If $n$ is (say) $10000$, then this is obviously too much to ask.

\subsection{Summary of Results\label{RESULTS}}

This paper initiates the study of the communication complexity and
computational complexity of agreement protocols. \ Its surprising conclusion
is that complexity is \textit{not} a major barrier to agreement---at least,
not nearly as major as it seems from the above arguments. \ In our view, this
conclusion strengthens Aumann's original theorem substantially, by forcing our
attention back to the origin of prior differences.

For economists, the main novelty of the paper will be our relentless use of
asymptotic analysis. \ We will never be satisfied to show that a protocol
terminates eventually. \ Instead we will always ask: do the resources needed
for the protocol scale `reasonably' with the parameters of the problem being
solved? \ Here `resources' include the number of messages, the number of bits
per message, and the number of computational steps; while `parameters' include
the number of agents, the number of bits each agent is given, and the desired
accuracy and probability of success. \ This approach will let us model the
limitations of real-world agents without sacrificing simplicity and elegance.

For computer scientists, the main novelty will be that, when we analyze the
communication complexity of a function $f$, we care only about how long it
takes some set of agents to agree \textit{among themselves} about the
expectation of $f$. \ Whether the agents' expectations agree with external
reality is irrelevant.

After introducing notation in Section \ref{PRELIM}, in Section \ref{CC}\ we
present our first set of results, which concern the communication complexity
of agreement.

Section \ref{UPPER}\ studies the \textquotedblleft economists' standard
protocol,\textquotedblright\ introduced by Geanakoplos and Polemarchakis
\cite{gp}\ and alluded to earlier. \ In that protocol, Alice and Bob
repeatedly announce their current expectations of a $\left[  0,1\right]
$\ random variable,\ conditioned on all previous announcements. \ The question
we ask is how many messages are needed before the agents' expectations agree
within $\varepsilon$ with probability at least $1-\delta$\ over their prior,
given parameters $\varepsilon$\ and $\delta$. \ We show that order $1/\left(
\delta\varepsilon^{2}\right)  $\ messages suffice. \ We then show that order
$1/\left(  \delta\varepsilon^{2}\right)  $\ messages still suffice, if instead
of sending their whole expectations (which are real numbers), the agents send
\textquotedblleft summary\textquotedblright\ messages consisting of only $2$
bits each. \ What makes these upper bounds surprising is that they are
completely independent of $n$, the number of bits needed to represent the
agents' knowledge. \ By contrast, in ordinary communication complexity (see
\cite{kn}), it is easy to show that given a random function $f:\left\{
0,1\right\}  ^{n}\times\left\{  0,1\right\}  ^{n}\rightarrow\left[
0,1\right]  $, Alice and Bob would need to exchange order $n$ bits to
approximate $f$ to within (say) $1/10$\ with high probability.

Given the results of Section \ref{UPPER}, several questions demand our
attention. \ Is the upper bound of $1/\left(  \delta\varepsilon^{2}\right)
$\ bits tight, or can it be improved even further?\ \ Also, is the economists'
standard protocol always optimal, or do other protocols sometimes need even
less communication? \ Section \ref{ATTEN} addresses these questions. \ Though
we are unable to show any lower bound better than $\log1/\varepsilon$ that
applies to \textit{all} protocols, we do give examples where the standard
protocol needs almost $1/\varepsilon^{2}$\ bits. \ We also show that the
standard protocol is not optimal: there exist cases where the standard
protocol uses almost $1/\varepsilon^{2}$\ bits, while a new protocol (which we
call the \textit{attenuated protocol}) uses fewer bits.

In earlier work, Parikh and Krasucki \cite{pk} extended Aumann's agreement
theorem to three or more agents, who send messages along the edges of a
directed graph.\ \ Thus, it is natural to ask whether our \textit{efficient
}agreement theorem extends to this setting as well. \ Section \ref{NAGENTS}
shows that it does: given $N$ agents with a common prior, who send messages
along a strongly connected graph of diameter $d$, order $Nd^{2}/\left(
\delta\varepsilon^{2}\right)  $\ messages suffice for every pair of agents to
agree within $\varepsilon$\ about the expectation of a $\left[  0,1\right]
$\ random variable with probability at least $1-\delta$\ over their prior.

In Section \ref{COMPUT} we shift attention to the \textit{computational}
complexity of agreement, the subject of our technically most interesting
result. \ What we want to show is that, even if two agents are computationally
bounded, after a conversation of reasonable length they can still probably
approximately agree about the expectation of a $\left[  0,1\right]  $\ random
variable. \ A large part of the problem is to say what this even means.
\ After all, if the agents both ignored their evidence and estimated (say)
$1/2$, then they would agree before exchanging a single message! \ So
agreement is only interesting if the agents have made some sort of
\textquotedblleft good-faith effort\textquotedblright\ to emulate Bayesian rationality.

Although we leave unspecified exactly what effort is necessary, we do propose
a criterion that we think is certainly \textit{sufficient}. \ This is that the
agents be able to \textit{simulate} a Bayesian agreement protocol, in such a
way that a computationally-unbounded referee, given the agents' knowledge
together with a transcript of their conversation, be unable to decide (with
non-negligible bias) whether the agents are computationally bounded or not.
\ The justification for this criterion is that, just as Turing \cite{turing}%
\ argued that a perfect simulation of thinking \textit{is} thinking, so it
seems to us that a statistically perfect simulation of Bayesian rationality
\textit{is} Bayesian rationality.

But what do we mean by computationally-bounded agents? \ We discuss this
question in detail in Section \ref{COMPUT}, but the basic point is that we
assume two \textquotedblleft subroutines\textquotedblright:\ one that computes
the $\left[  0,1\right]  $\ variable of interest, given a state of the world
$\omega$; and another that samples a state $\omega$ from any set in either
agent's initial knowledge partition. \ The complexity of the simulation
procedure is then expressed in terms of the number of calls to these subroutines.

Unfortunately, there is no way to simulate the economists' standard
protocol---even our discretized version of it---using a small number of
subroutine calls. \ The reason is that Alice's ideal estimate $p$ might lie on
a \textquotedblleft knife-edge\textquotedblright\ between the set of estimates
that would cause her to send message $m_{1}$ to Bob, and the set that would
cause her to send a different message $m_{2}$. \ In that case, it does not
suffice for her to approximate $p$ using random sampling; she needs to
determine it exactly. \ Our solution, which we develop in Section
\ref{SMOOTH}, is to have the agents \textquotedblleft smooth\textquotedblright%
\ their messages by adding random noise to them. \ By hiding small errors in
the agents' estimates, such noise makes the knife-edge problem disappear. \ On
the other hand, we show that in the computationally-unbounded case, the noise
does not prevent the agents from agreeing within $\varepsilon$\ with
probability $1-\delta$ after order $1/\left(  \delta\varepsilon^{2}\right)
$\ messages. \ In Sections \ref{SIMSMOOTH}\ and \ref{ANALYSIS}\ we prove the
main result: that the smoothed standard protocol can be simulated using a
number of subroutine calls that depends only on $\varepsilon$\ and $\delta$,
not on $n$. \ The dependence, alas, is exponential in $1/\left(  \delta
^{3}\varepsilon^{6}\right)  $, so our simulation procedure is still not
practical. \ However, we expect that both the procedure and its analysis can
be considerably improved.

We conclude in Section \ref{OPEN} with some suggestions for future research,
and some speculations about the causes of disagreement.

\section{Preliminaries\label{PRELIM}}

Let $\Omega$\ be a set of possible states of the world. \ Throughout this
paper, $\Omega$\ will be finite---both for simplicity of presentation, and
because we do not believe that any physically realistic agent can ever have
more than finitely many possible experiences. \ Let $\mathcal{D}$ be a prior
probability distribution over $\Omega$\ that is shared by some set of agents.
\ We can assume $\mathcal{D}$\ assigns nonzero probability to every $\omega
\in\Omega$, for if not, we simply remove the probability-$0$ states from
$\Omega$. \ Whenever we talk about a probability or expectation over a subset
$S$ of $\Omega$, unless otherwise indicated we mean that we start from
$\mathcal{D}$\ and conditionalize on $\omega\in S$.

Throughout this paper, we will consider protocols in which agents send
messages to each other in some order. \ Let $\Omega_{i,t}\left(
\omega\right)  $\ be the set of states that agent $i$ considers possible
immediately after the $t^{th}$\ message has been sent, given that the true
state of the world is $\omega$.\footnote{We assume for now that messages are
\textquotedblleft noise-free\textquotedblright; that is, they partition the
state space sharply. \ Later we will remove this assumption.} \ Then
$\omega\in\Omega_{i,t}\left(  \omega\right)  \subseteq\Omega$,\ and indeed the
set $\left\{  \Omega_{i,t}\left(  \omega\right)  \right\}  _{\omega\in\Omega}%
$\ forms a partition of $\Omega$. \ Furthermore, since the agents never forget
messages, we have $\Omega_{i,t}\left(  \omega\right)  \subseteq\Omega
_{i,t-1}\left(  \omega\right)  $. \ Thus we say that the partition $\left\{
\Omega_{i,t}\right\}  _{\omega\in\Omega}$\ \textit{refines} $\left\{
\Omega_{i,t-1}\right\}  _{\omega\in\Omega}$, or equivalently that $\left\{
\Omega_{i,t-1}\right\}  _{\omega\in\Omega}$\ \textit{coarsens} $\left\{
\Omega_{i,t}\right\}  _{\omega\in\Omega}$. \ (As a convention, we freely omit
arguments of $\omega$ when doing so will cause no confusion.) \ Notice also
that if the $t^{th}$\ message is not sent to agent $i$, then $\Omega
_{i,t}\left(  \omega\right)  =\Omega_{i,t-1}\left(  \omega\right)  $.

Now let $f:\Omega\rightarrow\left[  0,1\right]  $\ be a real-valued function
that the agents are interested in estimating.\ \ The assumption $f\left(
\omega\right)  \in\left[  0,1\right]  $\ is without loss of generality---for
since $\Omega$\ is finite, any function from $\Omega$ to $\mathbb{R}$\ has a
bounded range, which we can take to be $\left[  0,1\right]  $\ by rescaling.
\ We can think of $f\left(  \omega\right)  $ as the probability of some future
event conditioned on $\omega$, but this is not necessary. \ Let $E_{i,t}%
\left(  \omega\right)  =\operatorname*{EX}_{\omega^{\prime}\in\Omega
_{i,t}\left(  \omega\right)  }\left[  f\left(  \omega^{\prime}\right)
\right]  $\ be agent $i$'s expectation of $f$ at step $t$, given that the true
state of the world is $\omega$. \ Also, let $\Theta_{i,t}\left(
\omega\right)  =\left\{  \omega^{\prime}:E_{i,t}\left(  \omega^{\prime
}\right)  =E_{i,t}\left(  \omega\right)  \right\}  $\ be the set of states for
which agent $i$'s expectation of $f$ equals $E_{i,t}\left(  \omega\right)
$.\ \ Then the partition $\left\{  \Theta_{i,t}\right\}  _{\omega\in\Omega}%
$\ coarsens $\left\{  \Omega_{i,t}\right\}  _{\omega\in\Omega}$, and
$E_{i,t}\left(  \omega\right)  =\operatorname*{EX}_{\omega^{\prime}\in
\Theta_{i,t}\left(  \omega\right)  }\left[  f\left(  \omega^{\prime}\right)
\right]  $.

The following simple but important fact is due to Hanson \cite{hanson}.

\begin{proposition}
[\cite{hanson}]\label{hansonprop}Suppose the partition $\left\{  \Omega
_{i,t}\right\}  _{\omega\in\Omega}$\ refines $\left\{  \Theta_{j,u}\right\}
_{\omega\in\Omega}$. \ Then%
\[
\operatorname*{EX}_{\omega^{\prime}\in\Omega_{j,u}\left(  \omega\right)
}\left[  E_{i,t}\left(  \omega^{\prime}\right)  \right]  =\operatorname*{EX}%
_{\omega^{\prime}\in\Theta_{j,u}\left(  \omega\right)  }\left[  E_{i,t}\left(
\omega^{\prime}\right)  \right]  =E_{j,u}\left(  \omega\right)
\]
for all $\omega\in\Omega$. \ As a consequence, an agent's expectation of its
future expectation of $f$ always equals its current expectation. \ As another
consequence, if Alice has just communicated her expectation of $f$ to Bob,
then Alice's expectation of Bob's expectation of $f$ equals Alice's expectation.
\end{proposition}

\begin{proof}
In each case, we are taking the expectation of $f$\ over a subset
$S\subseteq\Omega$\ (either $\Omega_{j,u}\left(  \omega\right)  $\ or
$\Theta_{j,u}\left(  \omega\right)  $) for which $\operatorname*{EX}%
_{\omega^{\prime}\in S}\left[  f\left(  \omega^{\prime}\right)  \right]
=E_{j,u}\left(  \omega\right)  $. \ How $S$ is \textquotedblleft sliced
up\textquotedblright\ has no effect on the result.
\end{proof}

Proposition \ref{hansonprop}\ already demonstrates a dramatic difference
between Bayesian agreement protocols and actual human conversations. \ Suppose
Alice and Bob are discussing whether useful quantum computers will be built by
the year 2050. \ Bob says that, in his opinion, the chance of this happening
is only 5\%. \ Alice says she disagrees: she thinks the chance is 90\%. \ How
much should Alice expect her reply to influence Bob's estimate? \ Should she
expect him to raise it to 10\%, or even 15\%, out of deference to his friend
Alice's judgment? \ According to Proposition \ref{hansonprop}, she should
expect him to raise it to 90\%! \ That is, depending on what else Bob knows,
his new estimate might be 85\% or 95\%, but its \textit{expectation} from
Alice's point of view is 90\%.

\subsection{Miscellany\label{MISC}}

Asymptotic notation is standard: $F\left(  n\right)  =O\left(  G\left(
n\right)  \right)  $\ means there exist positive constants $a,b$ such that
$F\left(  n\right)  \leq a+bG\left(  n\right)  $\ for all $n\geq0$; $F\left(
n\right)  =\Omega\left(  G\left(  n\right)  \right)  $\ means the same but
with $F\left(  n\right)  \geq a+bG\left(  n\right)  $; $F\left(  n\right)
=\Theta\left(  G\left(  n\right)  \right)  $\ means $F\left(  n\right)
=O\left(  G\left(  n\right)  \right)  $\ and $F\left(  n\right)
=\Omega\left(  G\left(  n\right)  \right)  $;\ and $F\left(  n\right)
=o\left(  G\left(  n\right)  \right)  $\ means $F\left(  n\right)  =O\left(
G\left(  n\right)  \right)  $\ and not $F\left(  n\right)  =\Omega\left(
G\left(  n\right)  \right)  $.

We will have several occasions to use the following well-known bound.

\begin{theorem}
[Chernoff, Hoeffding]\label{hoeffding}Let $x_{1},\ldots,x_{K}$\ be $K$
independent samples of a $\left[  0,1\right]  $\ random variable with mean
$\mu$. \ Then for all $\alpha\in\left(  0,1\right)  $,%
\begin{align*}
\Pr\left[  x_{1}+\cdots+x_{K}\leq\left(  1-\alpha\right)  \mu K\right]   &
\leq e^{-\mu\alpha^{2}K/2},\\
\Pr\left[  \left\vert x_{1}+\cdots+x_{K}-\mu K\right\vert >\alpha K\right]
&  \leq2e^{-2\alpha^{2}K}.
\end{align*}

\end{theorem}

\section{Communication Complexity\label{CC}}

We now introduce and justify the communication complexity model. \ Assume for
the moment that there are two agents, Alice ($A$) and Bob ($B$); Section
\ref{NAGENTS}\ will generalize the model to three or more agents. \ We can
imagine if we like that Alice and Bob are given $n$-bit strings $x$ and $y$
respectively, so that $\Omega\subseteq\left\{  0,1\right\}  ^{n}\times\left\{
0,1\right\}  ^{n}$. \ Letting $\omega=\left(  x,y\right)  $, we then have
$\Omega_{A,0}\left(  \omega\right)  \subseteq x\times\left\{  0,1\right\}
^{n}$ and $\Omega_{B,0}\left(  \omega\right)  \subseteq\left\{  0,1\right\}
^{n}\times y$.

In an \textit{agreement protocol}, Alice and Bob take turns sending messages
to each other. \ Any such protocol is characterized by a sequence of functions
$m_{1},m_{2},\ldots:2^{\Omega}\rightarrow\mathcal{M}$, known to both agents,
which map subsets of $\Omega$\ to elements of a message space $\mathcal{M}$.
\ Possibilities for $\mathcal{M}$\ include $\left[  0,1\right]  $\ in a
continuous protocol, or\ $\left\{  0,1\right\}  $\ in a discretized protocol.
\ In all protocols considered in this paper, the $m_{t}$'s\ will be extremely
simple; for example, we might have $m_{t}\left(  S\right)  =\operatorname*{EX}%
_{\omega^{\prime}\in S}\left[  f\left(  \omega^{\prime}\right)  \right]  $ be
the agent's current expectation of $f$.

\begin{figure}
[ptb]
\begin{center}
\includegraphics[
trim=0.193566in 5.674407in 0.195636in 0.589387in,
height=0.7621in,
width=4.9365in
]%
{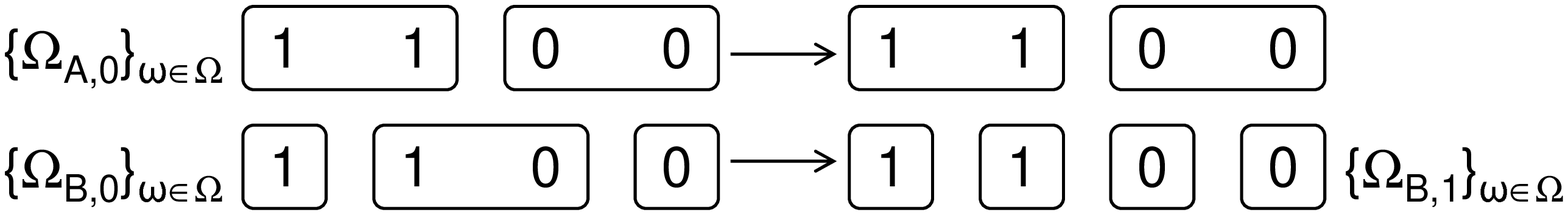}%
\caption{After Alice tells Bob whether $E_{A,0}$\ is $1$ or $0$, Bob's
partition $\left\{  \Omega_{B,0}\right\}  _{\omega\in\Omega}$\ is refined to
$\left\{  \Omega_{B,1}\right\}  _{\omega\in\Omega}$.}%
\label{partitionfig}%
\end{center}
\end{figure}

The protocol proceeds as follows: first Alice computes $m_{1}\left(
\Omega_{A,0}\left(  \omega\right)  \right)  $\ and sends it to Bob. \ After
seeing Alice's message, and assuming the true state of the world is $\omega$,
Bob's new set of possible states becomes%
\[
\Omega_{B,1}\left(  \omega\right)  =\Omega_{B,0}\left(  \omega\right)
\cap\left\{  \omega^{\prime}:m_{1}\left(  \Omega_{A,0}\left(  \omega^{\prime
}\right)  \right)  =m_{1}\left(  \Omega_{A,0}\left(  \omega\right)  \right)
\right\}
\]
as in Figure \ref{partitionfig}. \ Then Bob computes $m_{2}\left(
\Omega_{B,1}\left(  \omega\right)  \right)  $\ and sends it to Alice,
whereupon Alice's set of possible states becomes%
\[
\Omega_{A,2}\left(  \omega\right)  =\Omega_{A,0}\left(  \omega\right)
\cap\left\{  \omega^{\prime}:m_{2}\left(  \Omega_{B,1}\left(  \omega^{\prime
}\right)  \right)  =m_{2}\left(  \Omega_{B,1}\left(  \omega\right)  \right)
\right\}  .
\]
Then Alice computes $m_{3}\left(  \Omega_{A,2}\left(  \omega\right)  \right)
$\ and sends it to Bob, and so on.%
%TCIMACRO{\FRAME{ftbpFU}{4.9365in}{0.7621in}{0pt}{\Qcb{After Alice tells Bob
%whether $E_{A,0}$\ is $1$ or $0$, Bob's partition $\left\{  \Omega
%_{B,0}\right\}  _{\omega\in\Omega}$\ is refined to $\left\{  \Omega
%_{B,1}\right\}  _{\omega\in\Omega}$.}}{\Qlb{partitionfig}}{partition.eps}%
%{\special{ language "Scientific Word";  type "GRAPHIC";
%maintain-aspect-ratio TRUE;  display "USEDEF";  valid_file "F";
%width 4.9365in;  height 0.7621in;  depth 0pt;  original-width 10.3511in;
%original-height 7.7551in;  cropleft "0.0187";  croptop "0.9240";
%cropright "0.9811";  cropbottom "0.7317";
%filename 'partition.eps';file-properties "XNPEU";}}}%
%BeginExpansion
%EndExpansion

At this point we should address an obvious question: how do Alice and Bob know
each other's initial partitions, $\left\{  \Omega_{A,0}\right\}  _{\omega
\in\Omega}$\ and $\left\{  \Omega_{B,0}\right\}  _{\omega\in\Omega}$? \ If the
agents do not know each other's partitions, then messages between them are
useless, since neither agent knows how to update its own partition based on
the other's messages. \ This question is not specific to our setting; it can
be asked about Aumann's original result as well as any of its extensions.
\ The solution in each case is that the state of the world $\omega\in\Omega
$\ \textit{includes the agents' mental states as part of it}. \ From this it
follows that every agent has a uniquely defined partition known to every other
agent. \ For suppose Alice calculates that if the state of the world is
$\omega$, then Bob's knowledge is $\Omega_{B,0}\left(  \omega\right)  $,
meaning that he knows (and knows only) that the state belongs to $\Omega
_{B,0}\left(  \omega\right)  $. \ Then for all $\omega^{\prime}\in\Omega
_{B,0}\left(  \omega\right)  $, she must calculate that if the state is
$\omega^{\prime}$, then Bob's knowledge is $\Omega_{B,0}\left(  \omega\right)
$ as well. \ Otherwise one of her calculations was mistaken.

The reader might object on the following grounds. \ Suppose Alice and Bob are
the only two agents, and let $\Omega^{\left(  0\right)  }$\ be the set of
possible states of the \textquotedblleft external\textquotedblright%
\ world---meaning everything except Alice and Bob. \ Next let $\Omega^{\left(
1\right)  }$\ be the set of possible states of the agents' knowledge regarding
$\Omega^{\left(  0\right)  }$, let $\Omega^{\left(  2\right)  }$\ be the set
of possible states of their knowledge regarding $\Omega^{\left(  1\right)  }%
$,\ and so on. \ Then $\Omega=\Omega^{\left(  0\right)  }\times\Omega^{\left(
1\right)  }\times\Omega^{\left(  2\right)  }\times\cdots$, which contradicts
the assumption that $\Omega$\ is finite. \ The obvious response is that, since
the agents' brains can store only finitely many bits, not all elements of
$\Omega^{\left(  0\right)  }\times\Omega^{\left(  1\right)  }\times
\Omega^{\left(  2\right)  }\times\cdots$\ are actually possible.

However, the above response is open to a different objection, related to the
diagonalization arguments of G\"{o}del and Turing. \ Suppose Alice's and Bob's
brains store $n$ bits each. \ Then in order to reason about the set of
possible states of their brains, wouldn't they need brains that store more
than $n$ bits? \ We leave this conundrum unresolved, confining ourselves to
the following three remarks. \ First, only a tiny portion\ of the agents'
brains is likely to be relevant to their topic of conversation, which means
\textquotedblleft plenty of room left over\textquotedblright\ for
metareasoning about knowledge. \ Second, by reducing the number of brain
states that the agents need to consider, our results in Section \ref{COMPUT}%
\ will lessen the force of the self-reference argument, though not eliminate
it. \ Third, the agents' \textquotedblleft knowledge
hierarchy\textquotedblright\ seems likely to collapse at a low level. \ That
is, Alice might have little idea what sort of evidence shaped Bob's opinions
about the external world. \ But Bob probably has some idea what sort of
evidence shaped Alice's opinions about Bob's opinions, and Alice probably has
a good idea what sort of evidence shaped Bob's opinions about Alice's opinions
about Bob's opinions (assuming Bob even \textit{has} nontrivial such
opinions). \ The more indirect the knowledge, the fewer the ways of obtaining it.

Let us return to explaining the communication complexity model. \ After the
$t^{th}$ message, we say Alice and Bob $\left(  \varepsilon,\delta\right)
$\textit{-agree}\ if their expectations of $f$ agree to within $\varepsilon
$\ with probability at least $1-\delta$; that is, if%
\[
\Pr_{\omega\in\mathcal{D}}\left[  \left\vert E_{A,t}\left(  \omega\right)
-E_{B,t}\left(  \omega\right)  \right\vert >\varepsilon\right]  \leq\delta.
\]
The goal will be to minimize the number of messages until the agents $\left(
\varepsilon,\delta\right)  $-agree.

In our view, $\left(  \varepsilon,\delta\right)  $-agreement is a much more
fundamental notion than exact agreement.\ \ For suppose $f$ represents the
probability that global warming, if left unchecked, will cause sea levels to
rise\ at least $30$ centimeters\ by the year 2100. \ If after an hour's
conversation, any two people could agree within $1/4$ about $f$\ with
probability at least $3/4$, then the world would be a remarkably different
place than it now is. \ That the agreement was inexact and uncertain would be
less significant than the fact that it occurred at all.

But why do we calculate the success probability over $\mathcal{D}$, and not
some other distribution? \ In other words, what if the agents' priors agree
with each other, but not with external reality? \ Unfortunately, in that case
it seems difficult to prove anything, since the \textquotedblleft
true\textquotedblright\ prior could be concentrated on a few states that the
agents consider vanishingly unlikely. \ Furthermore, we conjecture that there
exist $f,\mathcal{D}$\ such that for all agreement protocols, the agents must
exchange $\Omega\left(  n\right)  $\ bits to agree within $\varepsilon$\ on
\textit{every} state $\omega$ (that is, to $\left(  \varepsilon,0\right)
$-agree). \ So given a protocol that causes Alice and Bob to $\left(
\varepsilon,\delta\right)  $-agree, what we should really say is that both
agents enter the conversation \textit{expecting} to agree within $\varepsilon
$\ with probability at least $1-\delta$. \ This, of course, is profoundly
unlike the situation in real life, where adversaries generally do not enter
arguments expecting to convince or to be convinced.

Let us make two further remarks about the model. \ First, if the agents want
to agree exactly (that is, $\left(  0,0\right)  $-agree), it is clear that in
the worst case they need $2n$ bits of communication, $n$ from Alice and $n$
from Bob. \ Note the contrast with ordinary communication complexity, where
$n$ bits always suffice. \ Indeed, even to produce approximate agreement,
two-way communication is necessary in general, as shown by the example
$f\left(  x,y\right)  =\left(  2x+y\right)  /3$, where $x,y\in\left\{
0,1\right\}  $ are uniformly distributed.

Second, our ending condition is simply that the agents $\left(  \varepsilon
,\delta\right)  $-agree at \textit{some} step $t$. \ We do not require them to
fix this $t$ independently of $f$ and $\mathcal{D}$. \ The reason is that for
any $t$, there might exist perverse $f,\mathcal{D}$\ such that the agents
nearly agree for the first $t-1$\ steps, then disagree violently at the
$t^{th}$\ step. \ However, it seems unfair to penalize the agents in such cases.

The following is the best lower bound we are able to show on agreement complexity.

\begin{proposition}
\label{logprop}There exist $f,\mathcal{D}$ such that for all $\varepsilon
\geq2^{-n}$ and $\delta\geq0$, Alice must send $\Omega\left(  \log
\frac{1-\delta}{\varepsilon}\right)  $\ bits to Bob and Bob must send
$\Omega\left(  \log\frac{1-\delta}{\varepsilon}\right)  $\ bits\ to Alice
before the agents $\left(  \varepsilon,\delta\right)  $-agree. \ In
particular, if $\delta$\ is bounded away from $1$ by a constant, then
$\Omega\left(  \log1/\varepsilon\right)  $\ bits are needed.
\end{proposition}

\begin{proof}
Let $\Omega=\left\{  1,\ldots,2^{n}\right\}  ^{2}$, let $\mathcal{D}$ be
uniform over $\Omega$, and let $f\left(  x,y\right)  =\left(  x+y\right)
/2^{n+1}$ for all $\left(  x,y\right)  \in\Omega$. \ Thus if $\widehat{x}$\ is
Bob's expectation of $x$ at step $t$ and $\widehat{y}$ is Alice's expectation
of $y$, then $E_{A,t}=\left(  x+\widehat{y}\right)  /2^{n+1}$\ and
$E_{B,t}=\left(  \widehat{x}+y\right)  /2^{n+1}$. \ Suppose one agent, say
Alice, has sent only $t<\log_{2}\left(  \frac{1-\delta}{\varepsilon}\right)
-2$\ bits to Bob. \ For each $i\in\left\{  1,\ldots,2^{t}\right\}  $, let
$p_{i}$\ be the probability of the $i^{th}$\ message sequence from Alice.
\ Conditioned on $i$, there are $2^{n}p_{i}$\ values of $x$\ still possible
from Bob's point of view.\ So regardless of $E_{B,t}$, the probability of
$\left\vert E_{A,t}-E_{B,t}\right\vert \leq\varepsilon$\ can be at most
$4\varepsilon/p_{i}$. \ Therefore the agents agree within $\varepsilon$\ with
total probability at most%
\[
\sum_{i=1}^{2^{t}}p_{i}\left(  \frac{4\varepsilon}{p_{i}}\right)
=4\varepsilon2^{t}<1-\delta.
\]

\end{proof}

\subsection{Convergence of the Standard Protocol\label{UPPER}}

The two-player \textquotedblleft standard protocol\textquotedblright\ is
simply the following: first Alice sends $E_{A,0}$, her current expectation of
$f$, to Bob.\ \ Then Bob sends his expectation $E_{B,1}$ to Alice, then Alice
sends $E_{A,2}$\ to Bob, and so on. \ Geanakoplos and Polemarchakis
\cite{gp}\ observed that for any $f,\mathcal{D}$, if the agents use the
standard protocol then after a finite number of messages $T$, they will reach
\textit{consensus}---meaning that $E_{A,T}=E_{B,T}$, both agents know this,
both know that they know it, etc. \ In particular, in our terminology Alice
and Bob $\left(  0,0\right)  $-agree.

In this section we ask how many messages are needed before the agents $\left(
\varepsilon,\delta\right)  $-agree. \ The surprising and unexpected answer, in
Theorem \ref{spthm}, is that $1/\left(  \delta\varepsilon^{2}\right)
$\ messages always suffice, independently of $n$ and all other parameters of
$f$\ and $\mathcal{D}$. \ One might guess that, since the expectations
$E_{A,0},E_{B,1},\ldots$\ are real numbers, the cost of communication must be
hidden in the length of the messages. \ However, in Theorem \ref{dspthm}\ we
show even if the agents send only $2$-bit \textquotedblleft
summaries\textquotedblright\ of their expectations, $O\left(  1/\left(
\delta\varepsilon^{2}\right)  \right)  $\ messages still suffice for $\left(
\varepsilon,\delta\right)  $-agreement.

Given any function $F:\Omega\rightarrow\left[  0,1\right]  $, let $\left\Vert
F\right\Vert _{2}^{2}=\operatorname*{EX}_{\omega\in\mathcal{D}}\left[
F\left(  \omega\right)  ^{2}\right]  $. \ The following proposition will be
used again and again in this paper.

\begin{proposition}
\label{refine}Suppose the partition $\left\{  \Omega_{i,t}\right\}
_{\omega\in\Omega}$\ refines $\left\{  \Theta_{j,u}\right\}  _{\omega\in
\Omega}$. \ Then%
\[
\left\Vert E_{i,t}\right\Vert _{2}^{2}-\left\Vert E_{j,u}\right\Vert _{2}%
^{2}=\left\Vert E_{i,t}-E_{j,u}\right\Vert _{2}^{2}%
\]
so in particular, $\left\Vert E_{i,t}\right\Vert _{2}^{2}\geq\left\Vert
E_{j,u}\right\Vert _{2}^{2}$. \ A special case\ is that $\left\Vert
E_{i,t+1}\right\Vert _{2}^{2}\geq\left\Vert E_{i,t}\right\Vert _{2}^{2}$ for
all $i,t$.
\end{proposition}

\begin{proof}
We have%
\[
\operatorname*{EX}\left[  E_{i,t}E_{j,u}\right]  =\operatorname*{EX}%
_{\omega\in\mathcal{D}}\left[  E_{j,u}\left(  \omega\right)
\operatorname*{EX}_{\omega^{\prime}\in\Theta_{j,u}\left(  \omega\right)
}\left[  E_{i,t}\left(  \omega^{\prime}\right)  \right]  \right]
=\operatorname*{EX}_{\omega\in\mathcal{D}}\left[  E_{j,u}\left(
\omega\right)  \cdot E_{j,u}\left(  \omega\right)  \right]  =\left\Vert
E_{j,u}\right\Vert _{2}^{2}%
\]
by Proposition \ref{hansonprop}, and therefore%
\[
\left\Vert E_{i,t}-E_{j,u}\right\Vert _{2}^{2}=\left\Vert E_{i,t}\right\Vert
_{2}^{2}+\left\Vert E_{j,u}\right\Vert _{2}^{2}-2\operatorname*{EX}\left[
E_{i,t}E_{j,u}\right]  =\left\Vert E_{i,t}\right\Vert _{2}^{2}-\left\Vert
E_{j,u}\right\Vert _{2}^{2}.
\]

\end{proof}

We can now prove an upper bound on the number of messages needed for agreement.

\begin{theorem}
\label{spthm}For all $f,\mathcal{D}$,\ the standard protocol causes Alice and
Bob to $\left(  \varepsilon,\delta\right)  $-agree after at most $1/\left(
\delta\varepsilon^{2}\right)  $ messages.
\end{theorem}

\begin{proof}
Intuitively, so long as the agents disagree by more than $\varepsilon$\ with
high probability, Alice's expectation $E_{A,1},E_{A,2},\ldots$ follows an
unbiased random walk with step size roughly $\varepsilon$. \ Furthermore, this
walk has two absorbing barriers at $0$ and $1$, for the simple fact that
$E_{A,t}\in\left[  0,1\right]  $. \ And we expect a random walk with step size
$\varepsilon$ to hit a barrier after about $1/\varepsilon^{2}$\ steps.

To make this intuition precise, we need only track the expectation, not of
$E_{A}$ and $E_{B}$, but of $E_{A}^{2}$ and $E_{B}^{2}$. \ Suppose Alice sends
the $t^{th}$\ message. \ Then Bob's partition $\left\{  \Omega_{B,t}\right\}
_{\omega\in\Omega}$ refines $\left\{  \Theta_{A,t-1}\right\}  _{\omega
\in\Omega}$. \ It follows by Proposition \ref{refine}\ that%
\[
\left\Vert E_{B,t}\right\Vert _{2}^{2}-\left\Vert E_{A,t-1}\right\Vert
_{2}^{2}=\left\Vert E_{B,t}-E_{A,t-1}\right\Vert _{2}^{2}.
\]
Assuming $\Pr\left[  \left\vert E_{B,t}-E_{A,t-1}\right\vert >\varepsilon
\right]  \geq\delta$, this implies that $\left\Vert E_{B,t}\right\Vert
_{2}^{2}>\left\Vert E_{A,t-1}\right\Vert _{2}^{2}+\delta\varepsilon^{2}$.
\ Similarly, after Bob sends Alice the $\left(  t+1\right)  ^{st}%
$\ message,\ we have $\left\Vert E_{A,t+1}\right\Vert _{2}^{2}>\left\Vert
E_{B,t}\right\Vert _{2}^{2}+\delta\varepsilon^{2}$. \ So until the agents
$\left(  \varepsilon,\delta\right)  $-agree, each message increases
$\max\left\{  \left\Vert E_{A,t}\right\Vert _{2}^{2},\left\Vert E_{B,t}%
\right\Vert _{2}^{2}\right\}  $\ by more than $\delta\varepsilon^{2}$. \ But
the maximum can never exceed $1$ (since $E_{A,t},E_{B,t}\in\left[  0,1\right]
$),\ which yields an upper bound of $1/\left(  \delta\varepsilon^{2}\right)  $
on the number of messages.
\end{proof}

As mentioned previously, the trouble with the standard protocol is that
sending one's expectation might require too many bits. \ A simple way to
discretize the protocol is as follows. \ Imagine a \textquotedblleft monkey in
the middle,\textquotedblright\ Charlie, who has the same prior distribution
$\mathcal{D}$ as Alice and Bob and who sees all messages between them, but who
does not know either of their inputs. \ In other words, letting $\Omega
_{C,t}\left(  \omega\right)  $\ be the set of states that Charlie considers
possible after the first $t$ messages, we have\ $\Omega_{C,0}\left(
\omega\right)  =\Omega$\ for all $\omega$.\ \ Then the partition $\left\{
\Omega_{C,t}\right\}  _{\omega\in\Omega}$\ coarsens both $\left\{
\Omega_{A,t}\right\}  _{\omega\in\Omega}$\ and $\left\{  \Omega_{B,t}\right\}
_{\omega\in\Omega}$;\ therefore both Alice and Bob can compute Charlie's
expectation $E_{C,t}\left(  \omega\right)  =\operatorname*{EX}_{\omega
^{\prime}\in\Omega_{C,t}\left(  \omega\right)  }\left[  f\left(
\omega^{\prime}\right)  \right]  $\ of $f$.

Now whenever it is her turn to send a message to Bob,\ Alice sends the message
\textquotedblleft high\textquotedblright\ if $E_{A,t}>E_{C,t}+\varepsilon/4$,
\textquotedblleft low\textquotedblright\ if $E_{A,t}<E_{C,t}-\varepsilon/4$,
and \textquotedblleft medium\textquotedblright\ otherwise. \ This requires $2$
bits. \ Likewise, Bob sends \textquotedblleft high\textquotedblright\ if
$E_{B,t}>E_{C,t}+\varepsilon/4$, \textquotedblleft low\textquotedblright\ if
$E_{B,t}<E_{C,t}-\varepsilon/4$, and \textquotedblleft
medium\textquotedblright\ otherwise.

\begin{theorem}
\label{dspthm}For all $f,\mathcal{D}$,\ the discretized protocol described
above causes Alice and Bob to $\left(  \varepsilon,\delta\right)  $-agree
after $O\left(  1/\left(  \delta\varepsilon^{2}\right)  \right)  $ messages.
\end{theorem}

\begin{proof}
The plan is to show that either $\left\Vert E_{A,t}\right\Vert _{2}^{2}$,
$\left\Vert E_{B,t}\right\Vert _{2}^{2}$, or $\left\Vert E_{C,t}\right\Vert
_{2}^{2}$\ increases by at least $\delta\varepsilon^{2}/512$\ with every
message of Alice's, until Alice and Bob $\left(  \varepsilon,\delta\right)
$-agree. \ Since $\left\Vert E_{i,t}\right\Vert _{2}^{2}\leq1$ for all $i$,
this will imply an upper bound of $3072/\left(  \delta\varepsilon^{2}\right)
$\ on the number of messages (we did not optimize the constant!).

Assume that $\Pr\left[  \left\vert E_{A,t}-E_{B,t}\right\vert >\varepsilon
\right]  \geq\delta$ and it is Alice's turn to send the $\left(  t+1\right)
^{st}$\ message.\ By the triangle inequality, either%
\[
\Pr\left[  \left\vert E_{A,t}-E_{C,t}\right\vert >\frac{\varepsilon}%
{2}\right]  \geq\frac{\delta}{2}%
\]
or%
\[
\Pr\left[  \left\vert E_{B,t}-E_{C,t}\right\vert >\frac{\varepsilon}%
{2}\right]  \geq\frac{\delta}{2}.
\]
We analyze these two cases separately. \ In the first case, with probability
at least $\delta/2$\ Alice's message is either \textquotedblleft
high\textquotedblright\ or \textquotedblleft low.\textquotedblright\ \ If the
message is \textquotedblleft high,\textquotedblright\ then $E_{C,t+1}%
$\ becomes an average\ of numbers\ each greater than $E_{C,t}+\varepsilon/4$,
so $E_{C,t+1}>E_{C,t}+\varepsilon/4$. \ If the message is \textquotedblleft
low,\textquotedblright\ then likewise $E_{C,t+1}<E_{C,t}-\varepsilon/4$.
\ Since $\left\{  \Omega_{C,t+1}\right\}  _{\omega\in\Omega}$\ refines
$\left\{  \Omega_{C,t}\right\}  _{\omega\in\Omega}$, Proposition
\ref{refine}\ thereby gives%
\[
\left\Vert E_{C,t+1}\right\Vert _{2}^{2}-\left\Vert E_{C,t}\right\Vert
_{2}^{2}=\left\Vert E_{C,t+1}-E_{C,t}\right\Vert _{2}^{2}>\frac{\delta}%
{2}\left(  \frac{\varepsilon}{4}\right)  ^{2}.
\]

Now for the second case. \ If, after Alice sends the $\left(  t+1\right)
^{st}$\ message, we still have%
\[
\Pr\left[  \left\vert E_{B,t+1}-E_{C,t+1}\right\vert >\frac{\varepsilon}%
{4}\right]  \geq\frac{\delta}{4},
\]
then the previous argument applied to Bob implies that%
\[
\left\Vert E_{C,t+2}\right\Vert _{2}^{2}-\left\Vert E_{C,t+1}\right\Vert
_{2}^{2}>\frac{\delta}{4}\left(  \frac{\varepsilon}{4}\right)  ^{2}%
\]
and we are done. \ So suppose otherwise. \ Then the difference between Bob's
and Charlie's expectations must have changed significantly:%
\[
\Pr\left[  \left\vert E_{B,t}-E_{C,t}\right\vert -\left\vert E_{B,t+1}%
-E_{C,t+1}\right\vert >\frac{\varepsilon}{4}\right]  >\frac{\delta}{4}.
\]
Hence by another application of the triangle inequality, either%
\[
\Pr\left[  \left\vert E_{B,t+1}-E_{B,t}\right\vert >\frac{\varepsilon}%
{8}\right]  >\frac{\delta}{8}%
\]
or%
\[
\Pr\left[  \left\vert E_{C,t+1}-E_{C,t}\right\vert >\frac{\varepsilon}%
{8}\right]  >\frac{\delta}{8}.
\]
In the former case, Proposition \ref{refine}\ yields%
\[
\left\Vert E_{B,t+1}\right\Vert _{2}^{2}-\left\Vert E_{B,t}\right\Vert
_{2}^{2}=\left\Vert E_{B,t+1}-E_{B,t}\right\Vert _{2}^{2}>\frac{\delta}%
{8}\left(  \frac{\varepsilon}{8}\right)  ^{2},
\]
while in the latter case,%
\[
\left\Vert E_{C,t+1}\right\Vert _{2}^{2}-\left\Vert E_{C,t}\right\Vert
_{2}^{2}>\frac{\delta}{8}\left(  \frac{\varepsilon}{8}\right)  ^{2}.
\]

\end{proof}

\subsection{Attenuated Protocol\label{ATTEN}}

We have seen that two agents, using the standard protocol, will always
$\left(  \varepsilon,\delta\right)  $-agree after exchanging only $O\left(
1/\left(  \delta\varepsilon^{2}\right)  \right)  $\ messages. \ This result
immediately raises three questions:

\begin{enumerate}
\item[(1)] Is there a scenario where the standard protocol \textit{needs}
about $1/\varepsilon^{2}$\ messages to produce $\left(  \varepsilon
,\delta\right)  $-agreement?

\item[(2)] Is the standard protocol always optimal, or do other protocols
sometimes outperform it?

\item[(3)] Is there a scenario where \textit{any} agreement protocol needs a
number of communication bits polynomial in $1/\varepsilon$?
\end{enumerate}

Although we leave question (3) open, in this section we resolve questions (1)
and (2). \ In particular, assume for simplicity that $\delta=1/2$. \ Then for
all $\varepsilon>0$, Theorem \ref{septhm} gives a scenario where the standard
protocol uses almost $1/\varepsilon^{2}$\ messages, even if the messages are
continuous rather discrete. \ By contrast, a new \textquotedblleft attenuated
protocol\textquotedblright\ uses only $2$ messages, both consisting of a
constant number of bits (independent of $\varepsilon$). \ Theorem
\ref{septhm2} then gives a \textit{fixed} scenario\ where for all
$\varepsilon>0$, the standard protocol uses almost $1/\varepsilon^{2}$
messages,\ while the attenuated protocol uses only $2$ messages, both
consisting of $O\left(  1/\varepsilon\right)  $ bits.

The attenuated protocol is interesting in its own right. \ The idea is to
imagine that in the standard protocol, the communication channel between Alice
and Bob becomes gradually more noisy as time goes on, so that each message
conveys slightly less information than the one before. \ It turns out that in
some cases, such noise would actually help! \ For intuitively, each time the
message intensity decreases by $\epsilon$, the \textquotedblleft
price\textquotedblright\ the agents pay in terms of disagreement is
proportional to $\epsilon^{2}$. \ So it is better for them to attenuate their
conversation gradually, than to send a sequence of \textquotedblleft
maximum-intensity\textquotedblright\ messages followed by no message (which we
can think of as intensity $0$).\footnote{The same phenomenon occurs in the
\textquotedblleft Zeno effect\textquotedblright\ of quantum mechanics .}
\ Even if the noise that produces this strange effect is missing from the
channel, the agents can easily simulate it. \ Furthermore, the messages will
turn out to be nonadaptive, so they can all be concatenated into one message
from Alice and one from Bob.

But how do we ensure that the standard protocol needs almost $1/\varepsilon
^{2}$\ messages? \ Intuitively, by forcing the random walk behavior of Section
\ref{UPPER}\ actually to occur. \ That is, at the beginning there will be a
disagreement that can only be resolved by Alice sending a bit to Bob. \ But
then that bit will cause a new disagreement even as it resolves the old one,
and so on.

\begin{theorem}
\label{septhm}For all $\varepsilon>0$,\ there exist $f,\mathcal{D}$\ such that
for all $\delta>0$:

\begin{enumerate}
\item[(i)] Using the standard protocol, Alice and Bob need to
exchange\ $\Omega\left(  \frac{1}{\varepsilon^{2}\log\frac{2}{\left(
1-\delta\right)  \varepsilon}}\right)  $\ messages before they $\left(
\varepsilon,\delta\right)  $-agree.

\item[(ii)] Using a different protocol, they need only exchange $2$ messages,
both consisting of $O\left(  \log1/\delta\right)  $ bits.
\end{enumerate}

In particular, if $\delta=1/2$ then the standard protocol needs $\Omega\left(
\frac{1/\varepsilon^{2}}{\log1/\varepsilon}\right)  $\ bits whereas the
attenuated protocol needs $O\left(  1\right)  $ bits.
\end{theorem}

\begin{proof}
Let%
\[
n=\frac{1}{64\varepsilon^{2}\ln\frac{6}{\left(  1-\delta\right)
\varepsilon^{2}}}%
\]
(throughout we omit floor and ceiling signs for convenience). \ The state
space $\Omega$ consists of all pairs $\left(  x,y\right)  $, where
$x=x_{1}\ldots x_{n}$ and $y=y_{1}\ldots y_{n}$\ belong to $\left\{
-1,1\right\}  ^{n}$. \ The prior distribution $\mathcal{D}$\ is uniform over
$\Omega$.\ \ Let%
\[
F\left(  x,y\right)  =\frac{1}{2}+2\varepsilon\sum_{i=1}^{n}\left(
y_{i-1}x_{i}+x_{i}y_{i}\right)
\]
where $y_{0}=1$. \ Then the function that interests the agents is%
\[
f\left(  x,y\right)  =\left\{
\begin{array}
[c]{cl}%
F\left(  x,y\right)  & \text{if }F\left(  x,y\right)  \in\left[  0,1\right] \\
0 & \text{if }F\left(  x,y\right)  <0\\
1 & \text{if }F\left(  x,y\right)  >1
\end{array}
\right.  .
\]
For simplicity, we first consider $F$ (which need not be bounded in $\left[
0,1\right]  $), and later analyze the \textquotedblleft edge
effects\textquotedblright\ that arise in switching to $f$. \ We claim that, if
the agents use the continuous standard protocol to evaluate $F$, then
$\left\vert E_{A,t}-E_{B,t}\right\vert =2\varepsilon$ at all steps $t<2n$,
where $E_{A,t}$ and $E_{B,t}$\ are Alice's and Bob's\ expectations of $F$
respectively after $t$ messages have been exchanged. \ For initially
$E_{A,0}=1/2+2\varepsilon x_{1}$\ and $E_{B,0}=1/2$. \ Most of the terms in
the sum defining $F\left(  x,y\right)  $ simply average to $0$ for both
agents, since Alice does not know the $y_{i}$'s and Bob does not know the
$x_{i}$'s. \ In the first step, however, the expectation that Alice sends to
Bob reveals $x_{1}$ to him. \ This causes $E_{B,1}$\ to become
$1/2+2\varepsilon x_{1}+2\varepsilon x_{1}y_{1}$, which differs from
$E_{A,0}=1/2+2\varepsilon x_{1}$ by $2\varepsilon$. \ Then in the second step,
the expectation that Bob sends to Alice reveals $y_{1}$\ to her, thereby
\textquotedblleft unlocking\textquotedblright\ the terms $x_{1}y_{1}$\ and
$y_{1}x_{2}$ in her expectation, and so on. \ It follows that until all
$2n$\ bits $x_{1}\ldots x_{n}$\ and $y_{1}\ldots y_{n}$\ have been exchanged,
the agents disagree by $2\varepsilon$ with certainty (see Figure
\ref{attenfig}).%
%TCIMACRO{\FRAME{ftbpFU}{3.3408in}{2.0872in}{0pt}{\Qcb{Alice's expectation
%$E_{A,t}$\ (solid line), and Bob's expectation $E_{B,t}$\ (dashed line), as a
%function of $t$}}{\Qlb{attenfig}}{atten.eps}%
%{\special{ language "Scientific Word";  type "GRAPHIC";
%maintain-aspect-ratio TRUE;  display "USEDEF";  valid_file "F";
%width 3.3408in;  height 2.0872in;  depth 0pt;  original-width 10.3511in;
%original-height 7.7551in;  cropleft "0.1443";  croptop "0.9879";
%cropright "0.8554";  cropbottom "0.3979";
%filename 'atten.eps';file-properties "XNPEU";}}}%
%BeginExpansion
\begin{figure}
[ptb]
\begin{center}
\includegraphics[
trim=1.493664in 3.085754in 1.496769in 0.093837in,
height=2.0872in,
width=3.3408in
]%
{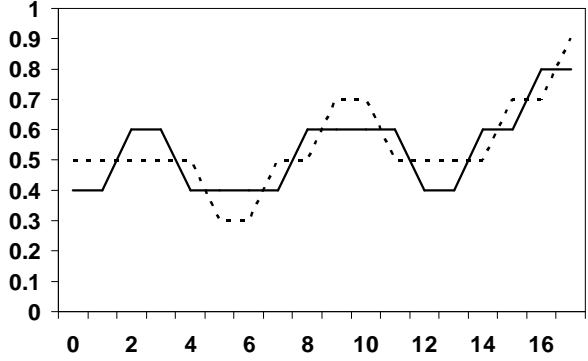}%
\caption{Alice's expectation $E_{A,t}$\ (solid line), and Bob's expectation
$E_{B,t}$\ (dashed line), as a function of $t$}%
\label{attenfig}%
\end{center}
\end{figure}
%EndExpansion

In switching from $F$ to $f$, the key observation is that Alice's expectation
$E_{A,t}\left(  f\right)  $\ of $f$ is a function of her expectation%
\[
E_{A,t}=\frac{1}{2}+2\varepsilon\left(  x_{1}+x_{1}y_{1}+y_{1}x_{2}%
+\cdots+x_{\left(  t-1\right)  /2}y_{\left(  t-1\right)  /2}+y_{\left(
t-1\right)  /2}x_{\left(  t+1\right)  /2}\right)
\]
of $F$. \ For from Alice's point of view, the later terms $x_{\left(
t+1\right)  /2}y_{\left(  t+1\right)  /2}$,\ $y_{\left(  t+1\right)
/2}x_{\left(  t+3\right)  /2}$, and so on are steps in an unbiased random walk
with starting point $E_{A,t}$, step size $2\varepsilon$, and \textquotedblleft
snapping barriers\textquotedblright\ at $0$ and $1$. \ (A snapping barrier is
neither absorbing nor reflecting: it allows a particle through, but if the
particle is found on the wrong side of the barrier after the walk ends, then
the particle is moved back to the barrier.) \ Let $E_{A,t}^{\ast}$\ be the
ending point of this walk; then $E_{A,t}\left(  f\right)  =\operatorname*{EX}%
\left[  E_{A,t}^{\ast}\right]  $ is a function of $E_{A,t}$. \ Likewise,
$E_{B,t}\left(  f\right)  =\operatorname*{EX}\left[  E_{B,t}^{\ast}\right]
$\ is the expected ending point of an unbiased walk with starting point
$E_{B,t}=E_{A,t}+2\varepsilon x_{\left(  t+1\right)  /2}y_{\left(  t+1\right)
/2}$, step size $2\varepsilon$, and snapping barriers at $0$ and $1$.

The lower bound for the standard protocol now follows from two claims:\ first,
that $E_{A,t}\in\left[  1/4,3/4\right]  $\ and $E_{B,t}\in\left[
1/4,3/4\right]  $\ for all $t\in\left\{  0,\ldots,2n\right\}  $ with
probability at least $\delta$. \ Second, that whenever $E_{A,t}$\ and
$E_{B,t}$ belong to $\left[  1/4,3/4\right]  $, we have $\left\vert
E_{A,t}\left(  f\right)  -E_{B,t}\left(  f\right)  \right\vert >\varepsilon$.
\ For the first claim, choose $z_{1},\ldots,z_{2n}$\ uniformly and
independently from $\left\{  0,1\right\}  $; then Theorem \ref{hoeffding}%
\ says that%
\[
\Pr\left[  \left\vert z_{1}+\cdots+z_{2n}-n\right\vert >\alpha\left(
2n\right)  \right]  \leq2e^{-4\alpha^{2}n}.
\]
Setting $\alpha=\frac{1/4}{2\varepsilon\left(  2n\right)  }$, this implies
that for any fixed $t$,%
\[
\Pr\left[  \left\vert E_{A,t}-1/4\right\vert >1/4\right]  \leq2e^{-1/\left(
64\varepsilon^{2}n\right)  }\leq\frac{1-\delta}{2n}%
\]
and similarly for $E_{B,t}$. \ The claim now follows from the union bound.
\ For the second claim, a bound similar to the above implies that%
\[
\Pr\left[  \left\vert E_{A,t}^{\ast}-E_{A,t}\right\vert >1/4\right]
\leq2e^{-1/\left(  64\varepsilon^{2}n\right)  }\leq\frac{\varepsilon}{3}%
\]
and similarly for $E_{B,t}^{\ast}$. \ This in turn implies that $\left\vert
E_{A,t}\left(  f\right)  -E_{A,t}\right\vert \leq\varepsilon/3$ and
$\left\vert E_{B,t}\left(  f\right)  -E_{B,t}\right\vert \leq\varepsilon/3$,
from whence it follows that $\left\vert E_{A,t}\left(  f\right)
-E_{B,t}\left(  f\right)  \right\vert >\varepsilon$ by the triangle inequality.

We now give the $O\left(  \log1/\delta\right)  $\ upper bound. \ It suffices
to give a protocol for $F$, since it is not hard to see that switching from
$F$ to $f$ can only decrease $\left\vert E_{A,t}-E_{B,t}\right\vert $. \ Let
$k=8\ln2/\delta$. \ For each $i\in\left\{  1,\ldots,k\right\}  $, Alice sends
Bob a bit that is uniformly random with probability $i/k$\ and $x_{i}%
$\ otherwise. \ Likewise, Bob sends Alice\ a bit that is uniformly random with
probability $i/k$\ and $y_{i}$\ otherwise. \ Then Alice's final expectation is%
\[
E_{A,2}=\frac{1}{2}+2\varepsilon\sum_{i=1}^{k}\left(  \frac{i-1}{k}%
y_{i-1}x_{i}+\frac{i}{k}x_{i}y_{i}\right)
\]
while Bob's is%
\[
E_{B,2}=\frac{1}{2}+2\varepsilon\sum_{i=1}^{k}\left(  \frac{i}{k}y_{i-1}%
x_{i}+\frac{i}{k}x_{i}y_{i}\right)  .
\]
So%
\[
E_{B,2}-E_{A,2}=\frac{2\varepsilon}{k}\sum_{i=1}^{k}y_{i-1}x_{i},
\]
and hence%
\[
\Pr\left[  \left\vert E_{A,2}-E_{B,2}\right\vert >\varepsilon\right]
=\Pr\left[  \left\vert z_{1}+\cdots+z_{k}-k/2\right\vert >k/4\right]
\]
where $z_{i}=\left(  y_{i-1}x_{i}+1\right)  /2$. \ Since the $z_{i}$'s are
uniform, independent samples from $\left\{  0,1\right\}  $, the above
probability is at most $2e^{-2\left(  1/4\right)  ^{2}k}=\delta$ by Theorem
\ref{hoeffding}.
\end{proof}

The main defect of Theorem \ref{septhm}\ is that the function $f$ had to be
tailored to a particular $\varepsilon$. \ The next theorem fixes this defect,
although the advantage of the attenuated protocol over the standard one is not
quite as dramatic as in Theorem \ref{septhm}. \ For simplicity,\ in stating
the theorem we fix $\delta=1/2$.

\begin{theorem}
\label{septhm2}For all $\gamma\in\left(  0,1\right)  $, there exist
$f$,$\mathcal{D}$ such that for all $\varepsilon\geq1/n^{1/\left(
2-\gamma\right)  }$:

\begin{enumerate}
\item[(i)] Using the standard protocol, Alice and Bob need to
exchange\ $\Omega\left(  1/\varepsilon^{2-\gamma}\right)  $\ messages before
they $\left(  \varepsilon,1/2\right)  $-agree.

\item[(ii)] Using the attenuated protocol, they need only exchange $2$
messages, both consisting of $O\left(  1/\varepsilon\right)  $ bits.
\end{enumerate}
\end{theorem}

\begin{proof}
[Sketch]Again we let $\mathcal{D}$\ be uniform over $x=x_{1}\ldots x_{n}$ and
$y=y_{1}\ldots y_{n}$\ in $\left\{  -1,1\right\}  ^{n}$. \ We then let%
\[
F\left(  x,y\right)  =\frac{1}{2}+\frac{\sqrt{\gamma}}{10}\sum_{i=1}^{n}%
\frac{y_{i-1}x_{i}+x_{i}y_{i}}{i^{1/\left(  2-\gamma\right)  }}%
\]
and%
\[
f\left(  x,y\right)  =\left\{
\begin{array}
[c]{cl}%
F\left(  x,y\right)  & \text{if }F\left(  x,y\right)  \in\left[  0,1\right] \\
0 & \text{if }F\left(  x,y\right)  <0\\
1 & \text{if }F\left(  x,y\right)  >1
\end{array}
\right.  .
\]
The rest of the proof is almost identical to that of Theorem \ref{septhm}, so
we omit it here.
\end{proof}

\subsection{$N$ Agents\label{NAGENTS}}

We have seen that two Bayesian agents can reach rapid agreement, provided they
communicate directly with each other. \ An obvious followup question is, what
if there are three or more agents, each of which talks only to its
`neighbors'? \ Will the agents still reach agreement, and if so, after how long?

Formally, let $G$ be a directed graph with vertices $1,\ldots,N$, each
representing an agent. \ Suppose messages can only be sent from agent $i$ to
agent $j$ if $\left(  i,j\right)  $\ is an edge in $G$. \ We need to assume
$G$ is strongly connected, since otherwise reaching agreement could be
impossible for trivial reasons. \ In this setting, a \textit{standard
protocol} consists of a sequence of edges $\left(  i_{1},j_{1}\right)
,\ldots,\left(  i_{t},j_{t}\right)  ,\ldots$ of $G$. \ At the $t^{th}$\ step,
agent $i_{t}$\ sends its current expectation $E_{i_{t},t-1}$\ of $f$ to agent
$j_{t}$, whereupon $j_{t}$\ updates its expectation accordingly. \ Call the
protocol \textit{fair} if every edge occurs infinitely often in the sequence.
\ Parikh and Krasucki \cite{pk}\ proved the following important theorem.

\begin{theorem}
[\cite{pk}]\label{pkthm}For all $f,\mathcal{D}$, any fair protocol will cause
all the agents' expectations to agree after a finite number of messages.
\end{theorem}

Indeed, the agents will reach \textit{consensus} after finitely many messages,
meaning it will be common knowledge among them that $E_{1,t}=\cdots=E_{N,t}$.
\ Here, though, we care only about the weaker condition of agreement.

\begin{figure}
[ptb]
\begin{center}
\includegraphics[
trim=3.710869in 5.081917in 3.523514in 0.098490in,
height=1.2951in,
width=1.565in
]%
{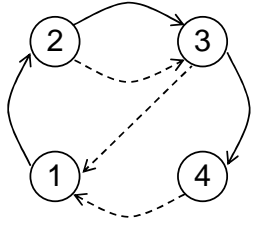}%
\caption{For a sample graph $G$, spanning tree $\mathcal{T}_{1}$\ is shown in
solid lines, and $\mathcal{T}_{2}$\ in dashed lines.}%
\label{nagentsfig}%
\end{center}
\end{figure}

Our goal is to cause every pair of agents to $\left(  \varepsilon
,\delta\right)  $-agree,\footnote{If we want every pair of agents to agree
within $\varepsilon$\ with \textit{global} probability $1-\delta$, then we
want every pair to $\left(  \varepsilon,\delta/N^{2}\right)  $-agree.} after a
number of steps polynomial in $N$, $1/\delta$, and $1/\varepsilon$. \ We can
achieve this via the following \textquotedblleft spanning-tree
protocol.\textquotedblright\ \ Let $\mathcal{T}_{1}$\ and $\mathcal{T}_{2}$ be
two spanning trees of $G$ of minimum diameter, both rooted at agent $1$. \ As
illustrated in Figure \ref{nagentsfig}, $\mathcal{T}_{1}$ points outward from
$1$ to the other $N-1$\ agents; $\mathcal{T}_{2}$ points inward back to $1$.
\ Let $\mathcal{O}_{1}$\ be an ordering of the edges of $\mathcal{T}_{1}$, in
which every edge originating at $i$ is preceded by an edge terminating at $i$,
unless $i=1$.\ \ Likewise let $\mathcal{O}_{2}$\ be an ordering of the edges
of $\mathcal{T}_{2}$, in which every edge originating at $i$ is preceded by an
edge terminating at $i$, unless $i$ is a leaf of $\mathcal{T}_{2}$. \ Then the
protocol is simply for agents to send their current expectations along edges
of $G$ in the order $\mathcal{O}_{1},\mathcal{O}_{2},\mathcal{O}%
_{1},\mathcal{O}_{2},\ldots$.%
%TCIMACRO{\FRAME{ftbpFU}{1.565in}{1.2951in}{0pt}{\Qcb{For a sample graph $G$,
%spanning tree $\mathcal{T}_{1}$\ is shown in solid lines, and $\mathcal{T}%
%_{2}$\ in dashed lines.}}{\Qlb{nagentsfig}}{nagents.eps}%
%{\special{ language "Scientific Word";  type "GRAPHIC";
%maintain-aspect-ratio TRUE;  display "USEDEF";  valid_file "F";
%width 1.565in;  height 1.2951in;  depth 0pt;  original-width 10.3511in;
%original-height 7.7551in;  cropleft "0.3585";  croptop "0.9873";
%cropright "0.6596";  cropbottom "0.6553";
%filename 'nagents.eps';file-properties "XNPEU";}}}%
%BeginExpansion
%EndExpansion

\begin{theorem}
\label{spthm3}For all $f,\mathcal{D}$,\ the spanning-tree protocol causes
every pair of agents to $\left(  \varepsilon,\delta\right)  $-agree after
$O\left(  \frac{Nd^{2}}{\delta\varepsilon^{2}}\right)  $ messages, where $d$
is the diameter of $G$.
\end{theorem}

\begin{proof}
We will track $\eta_{t}=\min_{i}\left\Vert E_{i,t}\right\Vert _{2}^{2}$.
\ Observe that, if the $t^{th}$\ message is from agent $i$ to agent $j$, then
the partition $\left\{  \Omega_{j,t+1}\right\}  _{\omega\in\Omega}$\ refines
both $\left\{  \Theta_{i,t}\right\}  _{\omega\in\Omega}$\ and $\left\{
\Omega_{j,t}\right\}  _{\omega\in\Omega}$, and therefore%
\[
\left\Vert E_{j,t+1}\right\Vert _{2}^{2}\geq\max\left\{  \left\Vert
E_{i,t}\right\Vert _{2}^{2},\left\Vert E_{j,t}\right\Vert _{2}^{2}\right\}
\]
by Proposition \ref{refine}. \ Also observe that, in any window of
$4N$\ messages, the spanning-tree protocol \textquotedblleft sends
information\textquotedblright\ from every agent to every other. \ Together
these observations imply that $\eta_{t+4N}\geq\max_{i}\left\Vert
E_{i,t}\right\Vert _{2}^{2}$. \ So as long as there exists an $i$\ such that
$\left\Vert E_{i,t}\right\Vert _{2}^{2}\gg\eta_{t}$, the protocol makes
significant progress.

It may happen, though, that $\left\Vert E_{i,t}\right\Vert _{2}^{2}$\ is
nearly constant as we range over $i$. \ Assume $\Pr\left[  \left\vert
E_{i,t}-E_{j,t}\right\vert >\varepsilon\right]  \geq\delta$\ for two agents
$i,j$. \ Consider a path from $i$ to $j$ in $G$, obtained by first following
$i$\ to $1$\ in $\mathcal{T}_{2}$\ and then following $1$\ to $j$ in
$\mathcal{T}_{1}$. \ This path has at most $2d$\ edges. \ So by the triangle
inequality, there exist consecutive agents $A,B$\ along the path such that%
\[
\left\Vert E_{A,t}-E_{B,t}\right\Vert _{2}\geq\frac{1}{2d}\left\Vert
E_{i,t}-E_{j,t}\right\Vert _{2}>\frac{\sqrt{\delta\varepsilon^{2}}}{2d}.
\]
Imagine that the $t^{th}$\ message is from $A$\ to $B$. \ Then since $\left\{
\Omega_{B,t+1}\right\}  _{\omega\in\Omega}$ refines both $\left\{
\Theta_{A,t}\right\}  _{\omega\in\Omega}$\ and $\left\{  \Omega_{B,t}\right\}
_{\omega\in\Omega}$, Proposition \ref{refine}\ yields%
\begin{align*}
\left\Vert E_{B,t+1}-E_{A,t}\right\Vert _{2}^{2}  &  =\left\Vert
E_{B,t+1}\right\Vert _{2}^{2}-\left\Vert E_{A,t}\right\Vert _{2}^{2},\\
\left\Vert E_{B,t+1}-E_{B,t}\right\Vert _{2}^{2}  &  =\left\Vert
E_{B,t+1}\right\Vert _{2}^{2}-\left\Vert E_{B,t}\right\Vert _{2}^{2}.
\end{align*}
Also, by the triangle inequality either%
\[
\left\Vert E_{B,t+1}-E_{A,t}\right\Vert _{2}^{2}\geq\frac{1}{4}\left\Vert
E_{A,t}-E_{B,t}\right\Vert _{2}^{2}%
\]
or%
\[
\left\Vert E_{B,t+1}-E_{B,t}\right\Vert _{2}^{2}\geq\frac{1}{4}\left\Vert
E_{A,t}-E_{B,t}\right\Vert _{2}^{2}.
\]
Therefore%
\[
\left\Vert E_{B,t+1}\right\Vert _{2}^{2}>\min\left\{  \left\Vert
E_{A,t}\right\Vert _{2}^{2},\left\Vert E_{B,t}\right\Vert _{2}^{2}\right\}
+\frac{1}{4}\left(  \frac{\delta\varepsilon^{2}}{4d^{2}}\right)  \geq\eta
_{t}+\frac{\delta\varepsilon^{2}}{16d^{2}}.
\]

It remains only to show why the above result is not spoiled if $A$\ or
$B$\ receive other messages before $A$\ sends its message to $B$. \ Let
$u$\ be the first time step after $t$\ in which $A$\ sends a message to $B$,
and suppose the steps between $t$\ and $u$ somehow reduce the distance between
$E_{A}$\ and $E_{B}$:%
\[
\left\Vert E_{A,u}-E_{B,u}\right\Vert _{2}^{2}\leq\frac{\delta\varepsilon^{2}%
}{16d^{2}}.
\]
Then by the triangle inequality (again!):%
\[
\left\Vert E_{A,u}-E_{A,t}\right\Vert _{2}+\left\Vert E_{B,u}-E_{B,t}%
\right\Vert _{2}\geq\left\Vert E_{B,t}-E_{A,t}\right\Vert _{2}-\left\Vert
E_{B,u}-E_{A,u}\right\Vert _{2}>\sqrt{\frac{\delta\varepsilon^{2}}{4d^{2}}%
}-\sqrt{\frac{\delta\varepsilon^{2}}{16d^{2}}}%
\]
so either%
\[
\left\Vert E_{A,u}-E_{A,t}\right\Vert _{2}^{2}>\frac{\delta\varepsilon^{2}%
}{64d^{2}}%
\]
or%
\[
\left\Vert E_{B,u}-E_{B,t}\right\Vert _{2}^{2}>\frac{\delta\varepsilon^{2}%
}{64d^{2}}.
\]
Suppose the former without loss of generality. \ Then since $\left\{
\Omega_{A,u}\right\}  _{\omega\in\Omega}$\ refines $\left\{  \Omega
_{A,t}\right\}  _{\omega\in\Omega}$,%
\[
\left\Vert E_{A,u}\right\Vert _{2}^{2}=\left\Vert E_{A,t}\right\Vert _{2}%
^{2}+\left\Vert E_{A,u}-E_{A,t}\right\Vert _{2}^{2}>\eta_{t}+\frac
{\delta\varepsilon^{2}}{64d^{2}}.
\]

We have shown that $\max_{i}\left\Vert E_{i,t+2N}\right\Vert _{2}^{2}=\eta
_{t}+\Omega\left(  \delta\varepsilon^{2}/d^{2}\right)  $, from which it
follows that $\eta_{t+6N}=\eta_{t}+\Omega\left(  \delta\varepsilon^{2}%
/d^{2}\right)  $. \ Hence the constraint $\eta_{t}\leq1$ yields an upper bound
of $O\left(  Nd^{2}/\left(  \delta\varepsilon^{2}\right)  \right)  $\ on the
number of messages.
\end{proof}

Let us make three remarks about Theorem \ref{spthm3}. \ First, naturally one
can combine Theorems \ref{spthm3} and \ref{dspthm}, to obtain an $N$-agent
protocol in which the messages are discrete. \ We omit the details here.
\ Second, all we really need about the \textit{order} of messages is that
information gets propagated from any agent in $G$ to any other in a reasonable
number of steps. \ Our spanning-tree construction was designed to guarantee
this, but sending messages in a random order (for example) would also work.
\ Third, it seems fair to assume that many agents send messages in parallel;
if so, our complexity bound can almost certainly be improved.

\section{Computational Complexity\label{COMPUT}}

The previous sections have weakened the idea that communication cost is a
fundamental barrier to agreement. \ However, we have glossed over the issue of
\textit{computational} cost entirely. \ A protocol that requires only
$O\left(  1/\left(  \delta\varepsilon^{2}\right)  \right)  $\ messages has
little real-world relevance if it would take Alice and Bob billions of years
to calculate the messages! \ Moreover, all protocols discussed above seem to
have that problem, since the number of possible states $\left\vert
\Omega\right\vert $\ could be exponential in the length $n$ of the agents' inputs.

Recognizing this issue, Hanson \cite{hanson3}\ introduced the notion of a
\textquotedblleft Bayesian wannabe\textquotedblright: a
computationally-bounded agent that can still make sense of what its
expectations would be if it had enough computational power to be a Bayesian.
\ He then showed that under certain assumptions, if two Bayesian wannabes
agree to disagree about the expectation of a function $f\left(  \omega\right)
$, then they must also disagree about some variable that is independent of the
state of the world $\omega\in\Omega$. \ However, Hanson's result does not
suggest a \textit{protocol} by which two Bayesian wannabes who agree about all
state-independent variables could come to agree about $f$\ as well.

Admittedly, if the two wannabes have \textit{very} limited abilities, it might
be trivial to get them to agree. \ For example, if Alice and Bob both ignore
all their evidence and estimate $f\left(  \omega\right)  =1/3$, then they
agree before exchanging even a single message. \ But this example seems
contrived: after all, if one the agents (with equal justification) estimated
$f\left(  \omega\right)  =2/3$, then no sequence of messages would ever cause
them to agree within $\varepsilon<1/3$. \ So informally, what we really want
to know is whether two wannabes will always agree, having put in a
\textquotedblleft good-faith effort\textquotedblright\ to emulate Bayesian rationality.

We are thus led to the following question. \ Is there an agreement protocol that

\begin{enumerate}
\item[(i)] would cause two computationally-unbounded Bayesians to $\left(
\varepsilon,\delta\right)  $-agree after a small number of messages, and

\item[(ii)] can be simulated using a small amount of computation?
\end{enumerate}

We will say shortly what we mean by a \textquotedblleft small amount of
computation.\textquotedblright\ \ By \textquotedblleft
simulate,\textquotedblright\ we mean that a computationally-unbounded referee,
given the state $\omega\in\Omega$ together with a transcript $M=\left(
m_{1},\ldots,m_{R}\right)  $\ of all messages exchanged during the protocol,
should be unable to decide (with non-negligible bias) whether Alice and Bob
were Bayesians following the protocol exactly, or Bayesian wannabes merely
simulating it. \ More formally, let $\mathcal{B}\left(  \omega\right)  $\ be
the probability distribution over message transcripts, assuming Alice and Bob
are Bayesians and the state of the world is $\omega$. \ Likewise, let
$\mathcal{W}\left(  \omega\right)  $\ be the distribution assuming Alice and
Bob are wannabes.\ \ Then we require that for all Boolean functions
$\Phi\left(  \omega,M\right)  $,%
\begin{equation}
\left\vert \Pr_{\omega\in\mathcal{D},M\in\mathcal{B}\left(  \omega\right)
}\left[  \Phi\left(  \omega,M\right)  =1\right]  -\Pr_{\omega\in
\mathcal{D},M\in\mathcal{W}\left(  \omega\right)  }\left[  \Phi\left(
\omega,M\right)  =1\right]  \right\vert \leq\zeta\tag{*}\label{star}%
\end{equation}
where $\zeta$\ is a parameter that can be made as small as we like (say
$0.00001$).

A consequence of the requirement (\ref{star}) is that even if Alice is
computationally unbounded, she cannot decide with bias greater than $\zeta$
whether Bob is also unbounded, judging only from the messages he sends to her.
\ For if Alice could decide, then so could our hypothetical referee, who
learns at least as much about Bob as Alice does. \ Though a little harder to
see, another consequence is that if Alice is unbounded, but knows Bob to be
bounded and \textit{takes his algorithm into account} when computing her
expectations, her messages will still be statistically indistinguishable from
what they would have been had she believed that Bob was unbounded. \ Indeed,
no beliefs, beliefs about beliefs, etc., about whether either agent is bounded
or not can significantly affect the sequence of messages, since the truth or
falsehood of those beliefs is almost irrelevant to predicting the agents'
future messages. \ Also, if Alice is unbounded for some steps of the protocol
but bounded for others, then Bob will never notice these changes, and would
hardly behave any differently were he told of them.

Because of these considerations, we claim that, while simulating a Bayesian
agreement protocol might not be the \textit{only} way for two Bayesian
wannabes to reach an \textquotedblleft honest\textquotedblright\ agreement, it
is certainly a \textit{sufficient} way. \ Therefore, if we can show how to
meet even the stringent requirement (\ref{star}), this will provide strong
evidence that computation time is not a fundamental barrier to agreement.

But what do we mean by computation time? \ We assume the state space $\Omega
$\ is a subset of $\left\{  0,1\right\}  ^{n}\times\left\{  0,1\right\}  ^{n}%
$, so that Alice's initial knowledge is an $n$-bit string $x$, and Bob's is an
$n$-bit string $y$. \ Given the prior distribution $\mathcal{D}$\ over
$\left(  x,y\right)  $ pairs, let $\mathcal{D}_{A,x}$\ be Alice's posterior
distribution over $y$ conditioned on $x$, and let $\mathcal{D}_{B,y}$ be Bob's
posterior distribution over $x$ conditioned on $y$. \ The following two
computational assumptions are the only ones that we make:

\begin{enumerate}
\item[(1)] Alice and Bob can both evaluate $f\left(  \omega\right)  $ for any
$\omega\in\Omega$.

\item[(2)] Alice and Bob can both sample from $\mathcal{D}_{A,x}$\ for any
$x\in\left\{  0,1\right\}  ^{n}$, and from $\mathcal{D}_{B,y}$\ for any
$y\in\left\{  0,1\right\}  ^{n}$.
\end{enumerate}

Our simulation procedure will \textit{not} have access to descriptions of $f$
or $\mathcal{D}$; it can learn about them only by calling subroutines for (1)
and (2) respectively. \ The complexity of the procedure will then be expressed
in terms of the number of subroutine calls, other computations adding a
negligible amount of time. \ Thus, we might stipulate that both subroutines
should run in time polynomial in $n$. \ On the other hand, $n$ could be
extremely large---otherwise the agents would simply exchange their entire
inputs and be done! \ So we probably want to be even stricter, and stipulate
that the subroutines should use time (say) \textit{logarithmic} in $n$, albeit
with many parallel processors. \ The latter seems like a better model for the
human brain; after all, to reach an opinion based on our current knowledge, we
do not contemplate every fact we know in sequential order, but instead zero in
quickly on the relevant facts. \ In any case, the simulation procedure will
treat the subroutines purely as \textquotedblleft black
boxes,\textquotedblright\ so decisions about their implementation will not
affect our results.

The justification for assumptions (1) and (2) is that without them, it is hard
to see how the agents could estimate their expectations even before they
started talking to each other. \ In other words, we have to assume the agents
enter the conversation with minimal tools for reasoning about their universe
of discourse. \ We do \textit{not} assume that those tools extend to reasoning
about each other's expectations, expectations of expectations, etc.,
conditioned on a sequence of messages exchanged. \ That the tools do extend in
this way is what we intend to prove.

The one assumption that seems debatable to us is that Alice can sample from
Bob's distribution $\mathcal{D}_{B,y}$, and Bob can sample from Alice's
distribution $\mathcal{D}_{A,y}$. \ How can an agent possibly be expected to
possess \textquotedblleft someone else's\textquotedblright\ sampling
subroutine? \ On further reflection, though, this question is simply a variant
of an earlier question: why can we assume that Alice knows Bob's set of
possible states $\Omega_{B}\left(  \omega\right)  $, and that Bob knows
$\Omega_{A}\left(  \omega\right)  $? \ For if Alice knows $\Omega_{B}\left(
\omega\right)  $\ as well as Bob does, then there is no particular reason why
she should not be able to sample from it as well as he can. \ Again, the
reason the agents know each other's partitions is that the state of the world
$\omega\in\Omega$ includes both agents' mental states as part of it. \ None of
this seems \textit{too} out of line with everyday experience---for whenever we
use what we know to try and figure out what someone else might be thinking, a
Bayesian would say we are sampling an $\omega$ from our set of possible
states, then sampling from what the other person's set of possible states
would be if the state of the world were $\omega$.

Finally, let us note that assumptions (1) and (2) can both be relaxed. \ In
particular, it is enough to approximate $f\left(  \omega\right)  $\ to within
an additive factor $\eta$ with probability at least $1-\eta$, in time that
increases polynomially in $1/\eta$. \ It is also enough to sample from a
distribution whose variation distance from $\mathcal{D}_{A,x}$\ or
$\mathcal{D}_{B,y}$\ is at most $\eta$, in time polynomial in $1/\eta$.
\ Indeed, since the probabilities and $f$-values are real numbers, we will
generally \textit{need} to approximate in order to represent them with finite
precision. \ For ease of presentation, though, we assume exact algorithms in
what follows.

\subsection{Smoothed Standard Protocol\label{SMOOTH}}

Na\"{\i}vely, requirement (\ref{star}) seems impossible to satisfy. \ All of
the agreement protocols discussed earlier in this paper---for example, that of
Theorem \ref{dspthm}---are easy to distinguish from any efficient simulation
of them. \ For consider Alice's first message to Bob. \ If
Alice's\ expectation $E_{A,0}$\ is below some threshold $c$, she sends one
message, whereas if $E_{A,0}\geq c$, she sends a different message. \ Even if
we fix $f$, and limit probabilities and $f$-values to (say) $n$ bits of
precision, we can arrange things so that $E_{A,0}\left(  \omega\right)  $\ is
exponentially close to $c$, sometimes greater and sometimes less, with high
probability over $\omega$. \ Then to decide which message to send, Alice needs
to evaluate $f$ exponentially many times.

We resolve this issue by having the agents add random noise to their messages
(\textquotedblleft smoothing\textquotedblright\ them), even if they are
unbounded Bayesians. \ This noise does not prevent the agents from reaching
$\left(  \varepsilon,\delta\right)  $-agreement. \ On the other hand, it makes
their messages easier to simulate. \ For unlike real numbers $a\neq b$, which
are perfectly distinguishable no matter how close they are, two probability
distributions with close means may be hard to distinguish, like wavepackets in
quantum mechanics.

\begin{figure}
[ptb]
\begin{center}
\includegraphics[
trim=0.780473in 4.888815in 0.391272in 0.000000in,
height=1.4371in,
width=4.5496in
]%
{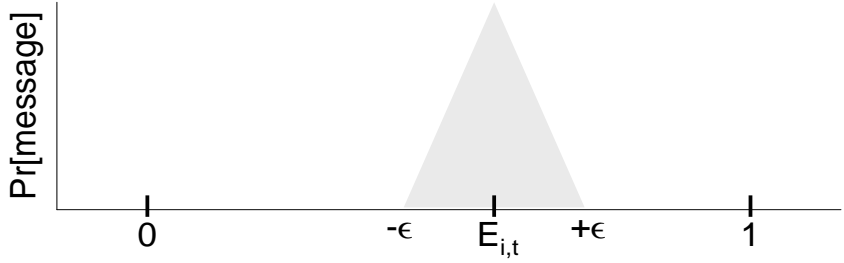}%
\caption{Agent $i$ \textquotedblleft smoothes\textquotedblright\ its
expectation $E_{i,t}$\ with triangular noise before sending it.}%
\label{triangular}%
\end{center}
\end{figure}

In the \textit{smoothed standard protocol}, Alice generates her messages to
Bob as follows. \ Let $b\geq\log_{2}\left(  200/\varepsilon\right)  $\ be a
positive integer to be specified later. \ Then let $\epsilon$ be an integer
multiple of $2^{-b}$\ between $\varepsilon/50$\ and $\varepsilon/40$, and let
$L=2^{b}\epsilon$. \ First Alice rounds her current expectation $E_{A,t}$\ of
$f$ to the nearest multiple of $2^{-b}$. \ Denote the result by
$\operatorname*{round}\left(  E_{A,t}\right)  $. \ She then draws an integer
$r\in\left\{  -L,\ldots,L\right\}  $, according to a \textit{triangular
distribution} in which $r=j$\ with probability $\left(  L-\left\vert
j\right\vert \right)  /L^{2}$ (see Figure \ref{triangular}). \ The message she
sends Bob is $m_{t+1}=\operatorname*{round}\left(  E_{A,t}\right)  +2^{-b}r$.
\ Observe that since $m_{t+1}\in\left[  -\epsilon,1+\epsilon\right]  $, there
are at most $2^{b}\left(  1+2\epsilon\right)  +1$\ possible values of
$m_{t+1}$---meaning Alice's message takes only $b+1$\ bits to specify. \ After
receiving the message, Bob updates his expectation of $f$ using Bayes' rule,
then draws an integer $r\in\left\{  -L,\ldots,L\right\}  $\ according to the
same triangular distribution and sends Alice $m_{t+2}=\operatorname*{round}%
\left(  E_{B,t+1}\right)  +2^{-b}r$. \ The two agents continue to send
messages in this way.%
%TCIMACRO{\FRAME{ftbpFU}{4.5496in}{1.4371in}{0pt}{\Qcb{Agent $i$
%\textquotedblleft smoothes\textquotedblright\ its expectation $E_{i,t}$\ with
%triangular noise before sending it.}}{\Qlb{triangular}}{triangular.eps}%
%{\special{ language "Scientific Word";  type "GRAPHIC";
%maintain-aspect-ratio TRUE;  display "USEDEF";  valid_file "F";
%width 4.5496in;  height 1.4371in;  depth 0pt;  original-width 10.3511in;
%original-height 7.7551in;  cropleft "0.0754";  croptop "1";
%cropright "0.9622";  cropbottom "0.6304";
%filename 'triangular.eps';file-properties "XNPEU";}}}%
%BeginExpansion
%EndExpansion

The reader might be wondering why we chose triangular noise, and whether other
types of noise would work equally well. \ The answer is that we want the
message distribution to have three basic properties. \ First, it should be
concentrated about a mean of $E_{i,t}$ with variance at most \symbol{126}%
$\epsilon^{2}$. \ Second, shifting the mean by $\eta\leq\epsilon$\ should
shift the distribution by at most \symbol{126}$\eta/\epsilon$\ in variation
distance. \ And third, the derivative of the probably density function should
never exceed \symbol{126}$\eta/\epsilon^{2}$\ in absolute value. \ Thus,
Gaussian noise would also work, though it is somewhat harder to analyze than
triangular noise. \ However, noise that is uniform over $\left[
-\epsilon,\epsilon\right]  $\ would \textit{not} work (so far as we could
tell), since it violates the third property.

Before we analyze the protocol, we need to develop some notation. \ Let
$M_{t}=\left(  m_{1},\ldots,m_{t}\right)  $\ consist of the first $t$ messages
that Alice and Bob exchange. \ Since messages are now probabilistic, the
agents' expectations of $f$ at step $t$ depend not only on the initial state
of the world $\omega$, but also on $M_{t}$. \ When we want to emphasize this,
we denote the agents' expectations by $E_{A,t}\left(  \omega,M_{t}\right)
$\ and $E_{B,t}\left(  \omega,M_{t}\right)  $\ respectively. \ Another
important consequence of messages being probabilistic is that after an agent
has received a message, its posterior distribution over $\omega$ is no longer
obtainable by restricting the prior distribution $\mathcal{D}$ to a subset\ of
possible states. \ Thus, we let $\Omega_{i}\left(  \omega\right)
=\Omega_{i,0}\left(  \omega\right)  $, since we will never refer to
$\Omega_{i,t}\left(  \omega\right)  $ for $t>0$.

Say the agents $\left(  \varepsilon,\delta\right)  $-agree\ after the $t^{th}%
$\ message if%
\[
\Pr_{\omega\in\Omega,M_{t}}\left[  \left\vert E_{A,t}\left(  \omega
,M_{t}\right)  -E_{B,t}\left(  \omega,M_{t}\right)  \right\vert >\varepsilon
\right]  \leq\delta.
\]
Also, let%
\[
\left\Vert E_{i,t}\right\Vert _{2}^{2}=\operatorname*{EX}_{\omega\in
\Omega,M_{t}}\left[  E_{i,t}\left(  \omega,M_{t}\right)  ^{2}\right]  .
\]

\begin{theorem}
\label{smoothed}For all $f,\mathcal{D}$, the smoothed standard protocol causes
Alice and Bob to $\left(  \varepsilon,\delta\right)  $-agree after at most
$2/\left(  \delta\varepsilon^{2}\right)  $\ messages.
\end{theorem}

\begin{proof}
Similarly to Theorem \ref{dspthm}, we let $E_{C,t}$\ be the expectation of a
third party Charlie who sees all messages between Alice and Bob, but who knows
neither their inputs nor\ the random bits that they use to produce their
messages. \ We then track $\left\Vert E_{C,t}\right\Vert _{2}^{2}$.

Assume that $\Pr\left[  \left\vert E_{A,t}-E_{B,t}\right\vert >\varepsilon
\right]  \geq\delta$ and that Alice sends the $t^{th}$\ message $m_{t}$.
\ Notice that $m_{t}$\ cannot deviate from Alice's expectation\ $E_{A,t}%
=E_{A,t-1}$\ by more than $2\epsilon$, since $\left\vert \operatorname*{round}%
\left(  E_{A,t}\right)  -E_{A,t}\right\vert \leq\epsilon$\ and $\left\vert
m_{t}-\operatorname*{round}\left(  E_{A,t}\right)  \right\vert \leq\epsilon$.
\ So keeping $M_{t}$\ fixed,%
\[
\left\vert E_{A,t}\left(  \omega,M_{t}\right)  -E_{A,t}\left(  \omega^{\prime
},M_{t}\right)  \right\vert \leq4\epsilon
\]
for all $\omega,\omega^{\prime}$. \ Now Charlie's expectation $E_{C,t}\left(
\omega,M_{t}\right)  $\ is just an average of $E_{A,t}\left(  \omega^{\prime
},M_{t}\right)  $'s, so it follows that%
\[
\left\vert E_{C,t}\left(  \omega,M_{t}\right)  -E_{A,t}\left(  \omega
,M_{t}\right)  \right\vert \leq4\epsilon
\]
as well. \ Similarly, after Bob sends the $\left(  t+1\right)  ^{st}%
$\ message,%
\[
\left\vert E_{C,t+1}\left(  \omega,M_{t+1}\right)  -E_{B,t+1}\left(
\omega,M_{t+1}\right)  \right\vert \leq4\epsilon.
\]
Therefore%
\[
\Pr\left[  \left\vert E_{C,t+1}-E_{C,t}\right\vert >\varepsilon-8\epsilon
\right]  \geq\Pr\left[  \left\vert E_{A,t}-E_{B,t}\right\vert >\varepsilon
\right]  \geq\delta,
\]
using the triangle inequality and the fact that $E_{B,t+1}=E_{B,t}$. \ The
final observation is that Charlie's partition of $\Omega\times M_{t+1}$\ at
step $t+1$\ refines his partition at step $t$, so by Proposition \ref{refine},%
\[
\left\Vert E_{C,t+1}\right\Vert _{2}^{2}-\left\Vert E_{C,t}\right\Vert
_{2}^{2}=\left\Vert E_{C,t+1}-E_{C,t}\right\Vert _{2}^{2}>\delta\left(
\varepsilon-8\epsilon\right)  ^{2}.
\]
Since $\left\Vert E_{C,t}\right\Vert _{2}^{2}\leq1$, this yields an upper
bound of $1/\left(  \delta\left(  \varepsilon-8\epsilon\right)  ^{2}\right)
<2/\left(  \delta\varepsilon^{2}\right)  $\ on the number of messages.
\end{proof}

\subsection{Simulating the Smoothed Protocol\label{SIMSMOOTH}}

Having proved that the smoothed standard protocol works, in this section we
explain how Alice and Bob can simulate the protocol. \ In the ideal
case---where the agents have unlimited computational power---they use the
following recursive formulas. \ Let%
\[
\Delta\left(  m_{t},E_{i,t-1}\right)  =\left\{
\begin{array}
[c]{cl}%
1-\left\vert m_{t}-\operatorname*{round}\left(  E_{i,t-1}\right)  \right\vert
/\epsilon & \text{if }\left\vert m_{t}-\operatorname*{round}\left(
E_{i,t-1}\right)  \right\vert \leq\epsilon\\
0 & \text{otherwise}%
\end{array}
\right.
\]
be proportional to the probability that agent $i$ sends message $m_{t}$, given
that its expectation is $E_{i,t-1}$. \ Also, let $q_{t}\left(  \omega
,M_{t}\right)  $\ be proportional to the joint probability of messages
$m_{1},\ldots,m_{t}$ assuming the true state of the world is $\omega$. \ Then
assuming $t$ is even and suppressing dependencies on $M_{t}$, for all
$X,Y$\ we have%
\begin{align*}
q_{t}\left(  Y\right)   &  =q_{t-2}\left(  Y\right)  \Delta\left(
m_{t},E_{B,t-1}\left(  Y\right)  \right)  ,\\
q_{t-1}\left(  X\right)   &  =q_{t-3}\left(  X\right)  \Delta\left(
m_{t-1},E_{A,t-2}\left(  X\right)  \right)  ,\\
E_{A,t}\left(  X\right)   &  =\frac{\operatorname*{EX}_{Y\in\Omega_{A}\left(
X\right)  }\left[  q_{t}\left(  Y\right)  f\left(  Y\right)  \right]
}{\operatorname*{EX}_{Y\in\Omega_{A}\left(  X\right)  }\left[  q_{t}\left(
Y\right)  \right]  },\\
E_{B,t-1}\left(  Y\right)   &  =\frac{\operatorname*{EX}_{X\in\Omega
_{B}\left(  Y\right)  }\left[  q_{t-1}\left(  X\right)  f\left(  X\right)
\right]  }{\operatorname*{EX}_{X\in\Omega_{B}\left(  Y\right)  }\left[
q_{t-1}\left(  X\right)  \right]  }%
\end{align*}
with the base cases $q_{0}\left(  Y\right)  =q_{-1}\left(  X\right)  =1$ for
all $X,Y$. \ The correctness of these formulas follows from simple Bayesian
manipulations. \ Having computed $E_{i,t}\left(  \omega\right)  $\ by the
formulas above (note that this does not require knowledge of $\omega$), all
agent $i$ needs to do is draw $r\in\left\{  -L,\ldots,L\right\}  $\ from the
triangular distribution, then send the message%
\[
m_{t+1}=\operatorname*{round}\left(  E_{i,t}\left(  \omega\right)  \right)
+2^{-b}r.
\]

In the real case, the agents are computationally bounded, and can no longer
afford the luxury of taking expectations over the exponentially large\ sets
$\Omega_{i}$. \ A natural idea is to compensate by somehow \textit{sampling}
those sets. \ But since we never assumed the ability to sample $\Omega_{i}%
$\ conditioned on messages $m_{1},\ldots,m_{t}$, it is not obvious how that
make that idea work. \ Our solution will consist of two phases: the
construction of \textquotedblleft sampling-trees,\textquotedblright\ which
involves no communication, followed by a message-by-message simulation of the
ideal protocol. \ Let us describe these phases in turn.

\textbf{(I) Sampling-Tree Construction.} \ Alice creates a tree $\mathcal{T}%
_{A}$\ with height $R$\ and branching factor $K$.\ \ Here $R<2/\left(
\delta\varepsilon^{2}\right)  $\ is the number of messages, and $K$ is a
parameter to be specified later. \ Let $\operatorname*{root}_{A}$\ be the root
node of $\mathcal{T}_{A}$, and let $S\left(  v\right)  $\ be the set of
children of node $v$. \ Then Alice labels each of the $K$\ nodes $w\in
S\left(  \operatorname*{root}_{A}\right)  $\ by a sample $Y_{w}\in\Omega
_{A}\left(  \omega\right)  $, drawn independently from her posterior
distribution $\mathcal{D}_{A,x}$. \ Next, for each $w\in S\left(
\operatorname*{root}_{A}\right)  $, she labels each of the $K$ nodes $v\in
S\left(  w\right)  $ by a sample $X_{v}\in\Omega_{B}\left(  Y_{w}\right)  $,
drawn independently from Bob's distribution $\mathcal{D}_{B,y}$ where
$Y_{w}=\left(  x,y\right)  $. \ She continues recursively in this manner,
labeling each $v$ an even distance from the root with a sample $X_{v}\in
\Omega_{B}\left(  Y_{w}\right)  $ where $w$ is the parent of $v$, and each $w$
an odd distance from the root with a sample $Y_{w}\in\Omega_{A}\left(
X_{v}\right)  $\ where $v$ is the parent of $w$. \ Thus her total number of
samples is%
\[
K+K^{2}+\cdots+K^{R}=\frac{K^{R+1}-1}{K-1}-1.
\]
Similarly, Bob creates a tree $\mathcal{T}_{B}$\ with height $R$\ and
branching factor $K$. \ Let $\operatorname*{root}_{B}$\ be the root of
$\mathcal{T}_{B}$; then Bob labels each $v\in S\left(  \operatorname*{root}%
_{B}\right)  $\ by a sample $X_{v}\in\Omega_{B}\left(  \omega\right)  $, each
child $w\in S\left(  v\right)  $\ of each $v\in S\left(  \operatorname*{root}%
_{B}\right)  $\ by a sample $Y_{w}\in\Omega_{A}\left(  X_{v}\right)  $, and so
on, alternating between $\Omega_{B}$\ and $\Omega_{A}$\ at successive levels.
\ As a side remark, if the agents share a random string, then there is no
reason for them not to use the same set of samples. \ However, we cannot
assume that such a string is available.

\textbf{(II) Simulation.} \ We now explain how the agents can use the samples
from (I) to simulate the smoothed standard protocol. \ First Alice estimates
her\ expectation $E_{A,0}$\ by the quantity%
\[
\left\langle E_{A,0}\left(  \operatorname*{root}\nolimits_{A}\right)
\right\rangle _{A}=\operatorname*{EX}_{w\in S\left(  \operatorname*{root}%
_{A}\right)  }\left[  f\left(  Y_{w}\right)  \right]  =\frac{1}{K}\sum_{w\in
S\left(  \operatorname*{root}_{A}\right)  }f\left(  Y_{w}\right)  .
\]
She then chooses a random $r\in\left\{  -L,\ldots,L\right\}  $ and sends Bob%
\[
m_{1}=\operatorname*{round}\left(  \left\langle E_{A,0}\left(
\operatorname*{root}\nolimits_{A}\right)  \right\rangle _{A}\right)
+2^{-b}r.
\]
On receiving the message, for each $v\in S\left(  \operatorname*{root}%
_{B}\right)  $\ Bob computes%
\[
\left\langle E_{A,0}\left(  v\right)  \right\rangle _{B}=\frac{1}{K}\sum_{w\in
S\left(  v\right)  }f\left(  Y_{w}\right)  ,
\]
his estimate of $E_{A,0}\left(  X_{v}\right)  $\ assuming $\omega=X_{v}%
$.\ \ He then defines%
\[
\left\langle q_{0}\left(  v\right)  \right\rangle _{B}=\Delta\left(
m_{1},\left\langle E_{A,0}\left(  v\right)  \right\rangle _{B}\right)
\]
and estimates his own expectation $E_{B,1}\left(  \omega\right)  $\ by%
\[
\left\langle E_{B,1}\left(  \operatorname*{root}\nolimits_{B}\right)
\right\rangle _{B}=\frac{\sum_{v\in S\left(  \operatorname*{root}_{B}\right)
}\left\langle q_{0}\left(  v\right)  \right\rangle _{B}f\left(  X_{v}\right)
}{\sum_{v\in S\left(  \operatorname*{root}_{B}\right)  }\left\langle
q_{0}\left(  v\right)  \right\rangle _{B}}.
\]
Finally, he chooses a random $r\in\left\{  -L,\ldots,L\right\}  $\ and sends
Alice%
\[
m_{2}=\operatorname*{round}\left(  \left\langle E_{B,1}\left(
\operatorname*{root}\nolimits_{B}\right)  \right\rangle _{B}\right)
+2^{-b}r.
\]
In general, if $t$ is even then the recursive formulas for agent $i$ are%

\begin{align*}
\left\langle q_{t}\left(  w\right)  \right\rangle _{i}  &  =\left\langle
q_{t-2}\left(  w\right)  \right\rangle _{i}\Delta\left(  m_{t},\left\langle
E_{B,t-1}\left(  w\right)  \right\rangle _{i}\right)  ,\\
\left\langle q_{t-1}\left(  v\right)  \right\rangle _{i}  &  =\left\langle
q_{t-3}\left(  v\right)  \right\rangle _{i}\Delta\left(  m_{t-1},\left\langle
E_{A,t-2}\left(  v\right)  \right\rangle _{i}\right)  ,\\
\left\langle E_{A,t}\left(  v\right)  \right\rangle _{i}  &  =\frac{\sum_{w\in
S\left(  v\right)  }\left\langle q_{t}\left(  w\right)  \right\rangle
_{i}f\left(  Y_{w}\right)  }{\sum_{w\in S\left(  v\right)  }\left\langle
q_{t}\left(  w\right)  \right\rangle _{i}},\\
\left\langle E_{B,t-1}\left(  w\right)  \right\rangle _{i}  &  =\frac
{\sum_{v\in S\left(  w\right)  }\left\langle q_{t-1}\left(  v\right)
\right\rangle _{i}f\left(  X_{v}\right)  }{\sum_{v\in S\left(  w\right)
}\left\langle q_{t-1}\left(  v\right)  \right\rangle _{i}}%
\end{align*}
with the base cases $\left\langle q_{0}\left(  w\right)  \right\rangle
_{i}=\left\langle q_{-1}\left(  v\right)  \right\rangle _{i}=1$ for all $w,v$.
\ Agent $i$ computes a message $m_{t}$\ in the obvious way, from its
expectation at the root of $\mathcal{T}_{i}$:%
\[
m_{t}=\operatorname*{round}\left(  \left\langle E_{i,t-1}\left(
\operatorname*{root}\nolimits_{i}\right)  \right\rangle _{i}\right)
+2^{-b}r.
\]

That completes the description of the simulation procedure. \ Its complexity
is easily determined: let $T_{1}$\ be the number of computational steps needed
to sample from $\mathcal{D}_{A,x}$\ or $\mathcal{D}_{B,y}$, and let $T_{2}%
$\ be number of steps needed to evaluate $f$. \ Then both agents use $O\left(
K^{R}\left(  T_{1}+T_{2}\right)  \right)  $\ steps, where we have summed over
all $R$ communication rounds. \ Thus, the complexity is exponential in
$R\approx2/\left(  \delta\varepsilon^{2}\right)  $; on the other hand, it has
no dependence on $n$.

\subsection{Analysis\label{ANALYSIS}}

Our goal is to show that the message sequence in the simulated protocol is
statistically indistinguishable from the sequence in the ideal protocol, for
some reasonable sample size $K$. \ Here `reasonable', unfortunately, is still
quite huge: of order $\left(  11/\epsilon\right)  ^{R^{2}}/\zeta^{2}$, where
$\zeta$\ is the maximum bias with which a referee can distinguish the
conversations. \ So assuming $\epsilon\geq\varepsilon/50$\ and $R\leq2/\left(
\delta\varepsilon^{2}\right)  $, the total number of computational steps is of
order%
\[
\left(  \frac{\left(  11/\epsilon\right)  ^{R^{2}}}{R^{2}}\right)  ^{R}\left(
T_{1}+T_{2}\right)  =\exp\left(  \frac{8\ln\left(  550/\varepsilon\right)
}{\delta^{3}\varepsilon^{6}}+\frac{4\ln\left(  1/\zeta\right)  }%
{\delta\varepsilon^{2}}\right)  \left(  T_{1}+T_{2}\right)  .
\]
The reader might complain that this bound is not at all reasonable: for
example, if $\varepsilon=\delta=1/2$, then it translates into more than
$2^{36864}$\ subroutine calls! \ Let us make two points in response. \ First,
we do show that the number of subroutine calls needed is independent of $n$,
and that it grows \textquotedblleft only\textquotedblright\ exponentially in a
polynomial in $1/\delta$\ and $1/\varepsilon$. \ Theoretical computer
scientists often see cases in which the first polynomial-time algorithm for a
problem has a completely impractical complexity, say $n^{40}$. \ However, once
the problem is known to be in polynomial time, it is usually possible to
reduce the exponent to obtain a truly practical algorithm. \ In our case, we
conjecture that the factor of $1/\left(  \delta^{3}\varepsilon^{6}\right)
$\ in the exponent could be reduced to $1/\left(  \delta^{2}\varepsilon
^{4}\right)  $\ or even $1/\left(  \delta\varepsilon^{2}\right)  $; certainly
the constants in the exponent can be reduced. \ The second point is that the
complexity is so large only because we never assumed the agents can sample
from their sets of possible states \textit{conditioned} on messages exchanged.
\ So the best they can do is to sample a huge number of states from their
original sets $\Omega_{A}$\ and $\Omega_{B}$, then retain the few that are
compatible with the messages. \ However, it seems likely that agents would
have at least some ability to sample conditioned on messages. \ After all, we
assumed that they enter the conversation with the ability to sample, and
presumably they have had other conversations in the past! \ In practice, then,
the complexity will probably be better than the worst-case estimate above.

How do we prove the simulation theorem? \ In one sense, the proof is `merely'
an exercise in error analysis and large deviation bounds. \ However, the
details are extremely subtle and difficult to get right. \ The problem is that
if a message has probability $q$ from its recipient's point of view, then
order $1/q$ samples are needed to find even a single input that could have
caused the sender to produce that message. \ Fortunately, low-probability
messages are unlikely to be sent, for almost tautological reasons that we
spell out in Lemma \ref{gamma}. \ However, because the sample trees
$\mathcal{T}_{i}$ are so large, with overwhelming probability they contain
\textit{some} nodes\ $v$\ with miniscule values of $\left\langle q_{t}\left(
v\right)  \right\rangle _{i}$. \ We need to argue that the errors introduced
by these \textquotedblleft bad nodes\textquotedblright\ are washed out by the
good nodes before they can propagate to the root.

The proof will repeatedly use the Chernoff-Hoeffding bound (Theorem
\ref{hoeffding}). \ As shown by the following corollary, Theorem
\ref{hoeffding}\ sometimes lets us estimate the mean of a random variable,
even if we cannot sample that variable directly.

\begin{corollary}
\label{chernoffcor}Let $p_{1},\ldots,p_{n}$\ and $x_{1},\ldots,x_{n}$\ belong
to $\left[  0,1\right]  $, and let $P=p_{1}+\cdots+p_{n}$ and $x=p_{1}%
x_{1}+\cdots+p_{n}x_{n}$. \ If we choose $K$\ indices $i\left(  1\right)
,\ldots,i\left(  K\right)  $\ uniformly at random from $\left\{
1,\ldots,n\right\}  $, then%
\[
\Pr\left[  \left\vert \frac{p_{i\left(  1\right)  }x_{i\left(  1\right)
}+\cdots+p_{i\left(  K\right)  }x_{i\left(  K\right)  }}{p_{i\left(  1\right)
}+\cdots+p_{i\left(  K\right)  }}-\frac{x}{P}\right\vert >\alpha\right]
\leq4e^{-\alpha^{2}\left(  P/n\right)  ^{2}K/2}.
\]

\end{corollary}

\begin{proof}
Let%
\begin{align*}
\widetilde{P}  &  =\frac{n}{K}\left(  p_{i\left(  1\right)  }+\cdots
+p_{i\left(  K\right)  }\right)  ,\\
\widetilde{X}  &  =\frac{n}{K}\left(  p_{i\left(  1\right)  }x_{i\left(
1\right)  }+\cdots+p_{i\left(  K\right)  }x_{i\left(  K\right)  }\right)  .
\end{align*}
Then since $\widetilde{X}\leq\widetilde{P}$,%
\[
\left\vert \frac{\widetilde{X}}{\widetilde{P}}-\frac{X}{P}\right\vert
=\frac{\left\vert \widetilde{X}\left(  P-\widetilde{P}\right)  -\widetilde
{P}\left(  X-\widetilde{X}\right)  \right\vert }{\widetilde{P}P}\leq
\frac{\left\vert \widetilde{P}-P\right\vert }{P}+\frac{\left\vert
\widetilde{X}-X\right\vert }{P}.
\]
So%
\[
\Pr\left[  \left\vert \frac{\widetilde{X}}{\widetilde{P}}-\frac{X}%
{P}\right\vert >\alpha\right]  \leq\Pr\left[  \left\vert \widetilde
{P}-P\right\vert >\frac{\alpha P}{2}\right]  +\Pr\left[  \left\vert
\widetilde{X}-X\right\vert >\frac{\alpha P}{2}\right]  .
\]
By Theorem \ref{hoeffding},%
\[
\Pr\left[  \frac{K}{n}\left\vert \widetilde{P}-P\right\vert >\frac{\alpha
P}{2n}K\right]  \leq2e^{-\alpha^{2}\left(  P/n\right)  ^{2}K/2}%
\]
and similarly for $\left\vert \widetilde{X}-X\right\vert $.
\end{proof}

We will also need a bound for a sum of exponentially distributed variables,
which can be found in \cite{dp}\ for example.

\begin{theorem}
\label{exponential}Let $x_{1},\ldots,x_{K}\in\left[  0,\infty\right)
$\ be\ independent and exponentially distributed with mean $1$ (that is,
$\Pr\left[  x_{i}\geq x\right]  =e^{-\omega}$). \ Then%
\[
\Pr\left[  x_{1}+\cdots+x_{K}\geq\left(  1+\alpha\right)  K\right]
\leq\left(  \frac{e^{\alpha}}{1+\alpha}\right)  ^{-K}.
\]

\end{theorem}

For convenience, we will state our results in terms of Alice's tree
$\mathcal{T}_{A}$, with the understanding that they apply equally well to
$\mathcal{T}_{B}$. \ Throughout, we assume that $t$\ is even and that the
$t^{th}$\ message $m_{t}$\ is sent from Bob to Alice. \ Let $Q_{t}=\sum
_{Y\in\Omega_{A}\left(  \omega\right)  }q_{t}\left(  Y\right)  $\ measure the
\textquotedblleft likelihood\textquotedblright\ of Alice's situation at step
$t$. \ Then $Q_{t}/Q_{t-2}$\ measures the likelihood of\ the $t^{th}%
$\ message, conditioned on Alice's situation just before she receives it.
\ The following lemma says essentially that \textquotedblleft unlikely
messages are unlikely.\textquotedblright

\begin{lemma}
\label{gamma}For all inputs $x$ of Alice, message sequences $M_{t-1}$, and
constants $\gamma>0$,%
\[
\Pr_{m_{t}}\left[  \frac{Q_{t}}{Q_{t-2}}\leq\frac{\gamma\epsilon}{2}\right]
<\gamma.
\]

\end{lemma}

\begin{proof}
For all $m\in\left[  -\epsilon,1+\epsilon\right]  $,%
\begin{align*}
\Pr\left[  m_{t}=m\right]   &  =\sum_{j\in\left\{  -L,\ldots,L\right\}
}\left(  \Pr_{Y}\left[  \operatorname*{round}\left(  E_{B,t-1}\left(
Y\right)  \right)  =m+2^{-b}j\right]  \cdot\frac{\Delta\left(  m,m+2^{-b}%
j\right)  }{L}\right) \\
&  =\frac{1}{L}\frac{\sum_{Y\in\Omega_{A}\left(  \omega\right)  }%
q_{t-2}\left(  Y\right)  \Delta\left(  m,E_{B,t-1}\left(  Y\right)  \right)
}{\sum_{Y\in\Omega_{A}\left(  \omega\right)  }q_{t-2}\left(  Y\right)  }%
=\frac{1}{L}\frac{Q_{t}}{Q_{t-2}}%
\end{align*}
from Alice's point of view. \ So it suffices to observe that%
\[
\Pr_{m}\left[  \Pr_{m_{t}}\left[  m_{t}=m\right]  \leq\frac{\gamma\epsilon
}{2L}\right]  \leq\frac{\gamma\epsilon}{2L}\frac{L\left(  1+2\epsilon\right)
+\epsilon}{\epsilon}<\gamma.
\]
Here the first inequality follows from elementary probability theory, together
with the fact that there are at most $\left(  1+2\epsilon\right)  /2^{-b}%
+1$\ possible messages $m$, and hence the mean of $\Pr_{m_{t}}\left[
m_{t}=m\right]  $\ over $m$ chosen uniformly at random is at least%
\[
\frac{1}{\left(  1+2\epsilon\right)  /2^{-b}+1}=\frac{\epsilon}{L\left(
1+2\epsilon\right)  +\epsilon}.
\]
The second inequality follows since $\epsilon<1/4$.
\end{proof}

A consequence of Lemma \ref{gamma}\ is that unlikely \textit{sequences} of
messages are unlikely. \ For the remainder of this section, let $g=\frac
{4e}{\epsilon}\ln K$.

\begin{lemma}
\label{stillbig}For all $\gamma>0$ and all $x$,%
\[
\Pr_{y,M_{t}}\left[  Q_{t}\leq\gamma\right]  <g^{t/2}\max\left\{  \gamma
,\frac{1}{K}\right\}  .
\]

\end{lemma}

\begin{proof}
For all $u\in\left\{  2,4,\ldots,t\right\}  $, let $x_{u}=\ln\left(  \epsilon
Q_{u-2}/2Q_{u}\right)  $. \ Then%
\[
Q_{t}=\frac{2Q_{t}}{\epsilon Q_{t-2}}\frac{2Q_{t-2}}{\epsilon Q_{t-4}}%
\cdots\frac{2Q_{2}}{\epsilon Q_{0}}\left(  \frac{\epsilon}{2}\right)
^{t/2}=e^{-x_{2}-x_{4}-\cdots-x_{t}}\left(  \frac{\epsilon}{2}\right)  ^{t/2}%
\]
since $Q_{0}=1$. \ Furthermore, Lemma \ref{gamma} implies that for each $u$,%
\[
\Pr\left[  x_{u}\geq x\right]  =\Pr_{m_{u}}\left[  \frac{Q_{u}}{Q_{u-2}}%
\leq\frac{e^{-x}\epsilon}{2}\right]  <e^{-x},
\]
even conditioned on $x_{2},\ldots,x_{u-2}$. \ Therefore $x_{2}+\cdots+x_{t}%
$\ is stochastically dominated by a sum of $t/2$\ independent exponential
variables each with mean $1$. \ So by Theorem \ref{exponential},%
\[
\Pr\left[  x_{2}+\cdots+x_{t}\geq\left(  1+\alpha\right)  \frac{t}{2}\right]
<\left(  \frac{e^{\alpha}}{1+\alpha}\right)  ^{-t/2}.
\]
Setting $\gamma=e^{-\left(  1+\alpha\right)  t/2}\left(  \epsilon/2\right)
^{t/2}$\ and solving to obtain $\alpha=\left(  2/t\right)  \ln\left(  \left(
\epsilon/2\right)  ^{t/2}/\gamma\right)  -1$, it follows that \
\[
\Pr_{y,M_{t}}\left[  Q_{t}\leq\gamma\right]  <\left(  \frac{e^{\left(
2/t\right)  \ln\left(  \left(  \epsilon/2\right)  ^{t/2}/\gamma\right)  -1}%
}{\left(  2/t\right)  \ln\left(  \left(  \epsilon/2\right)  ^{t/2}%
/\gamma\right)  }\right)  ^{-t/2}<\left(  \frac{4e}{\epsilon}\ln\frac
{1}{\gamma}\right)  ^{t/2}\gamma\leq g^{t/2}\max\left\{  \gamma,\frac{1}%
{K}\right\}  .
\]

\end{proof}

In the next four results, we fix a particular node $v\in\mathcal{T}_{A}$, then
study how the error at $v$ depends on the errors at its children $w\in
S\left(  v\right)  $. \ For simplicity, we assume $v$ is an even distance from
the root, but our results will apply equally to nodes an odd distance from the
root. \ We need to upper-bound the expected difference between Alice's actual
expectation $\left\langle E_{A,t}\left(  v\right)  \right\rangle _{A}$, and
her ideal expectation $E_{A,t}\left(  X_{v}\right)  $. \ To this end, it will
be helpful to define the following \textquotedblleft hybrid\textquotedblright%
\ between $\left\langle E_{A,t}\left(  v\right)  \right\rangle _{A}$ and
$E_{A,t}\left(  X_{v}\right)  $:%
\[
E_{A,t}^{\ast}\left(  v\right)  =\frac{\sum_{w\in S\left(  v\right)  }%
q_{t}\left(  Y_{w}\right)  f\left(  Y_{w}\right)  }{\sum_{w\in S\left(
v\right)  }q_{t}\left(  Y_{w}\right)  }.
\]
To compute $E_{A,t}^{\ast}$, we use the ideal weights $q_{t}\left(
Y_{w}\right)  $, but we average over Alice's $K$ samples $\left\{
Y_{w}\right\}  _{w\in S\left(  v\right)  }$ only, not over all of $\Omega
_{A}\left(  X_{v}\right)  $. \ By the triangle inequality, to upper-bound
$\left\vert \left\langle E_{A,t}\left(  v\right)  \right\rangle _{A}%
-E_{A,t}\left(  X_{v}\right)  \right\vert $\ it suffices to upper-bound
$\left\vert \left\langle E_{A,t}\left(  v\right)  \right\rangle _{A}%
-E_{A,t}^{\ast}\left(  v\right)  \right\vert $\ and $\left\vert E_{A,t}^{\ast
}\left(  v\right)  -E_{A,t}\left(  X_{v}\right)  \right\vert $. \ We start
with the latter.

\begin{lemma}
\label{integrals}%
\[
\operatorname*{EX}_{y,M_{t},S\left(  v\right)  }\left[  \left\vert
E_{A,t}^{\ast}\left(  v\right)  -E_{A,t}\left(  X_{v}\right)  \right\vert
\right]  \leq\frac{7g^{t/2+1}}{\sqrt{K}}.
\]

\end{lemma}

\begin{proof}
Assuming $Q_{t}=Q$,%
\[
\Pr\left[  \left\vert E_{A,t}^{\ast}\left(  v\right)  -E_{A,t}\left(
X_{v}\right)  \right\vert \geq\omega\right]  \leq4e^{-\omega^{2}Q^{2}K/2}%
\]
by Corollary \ref{chernoffcor}. \ Furthermore, since $E_{A,t}^{\ast}\left(
v\right)  $\ and $E_{A,t}\left(  X_{v}\right)  $\ are in $\left[  0,1\right]
$, we have the trivial but important bound $\left\vert E_{A,t}^{\ast}\left(
v\right)  -E_{A,t}\left(  X_{v}\right)  \right\vert \leq1$. \ Therefore%
\begin{align*}
\operatorname*{EX}\left[  \left\vert E_{A,t}^{\ast}\left(  v\right)
-E_{A,t}\left(  X_{v}\right)  \right\vert \right]   &  =\int_{0}^{1}\Pr\left[
\left\vert E_{A,t}^{\ast}\left(  v\right)  -E_{A,t}\left(  X_{v}\right)
\right\vert \geq x\right]  dx\\
&  \leq4\int_{0}^{1}\operatorname*{EX}_{Q_{t}}\left[  e^{-x^{2}Q_{t}^{2}%
K/2}\right]  dx\\
&  =4\int_{0}^{1}\int_{0}^{1}\Pr\left[  e^{-x^{2}Q_{t}^{2}K/2}\geq x\right]
dxdx\\
&  =4\int_{0}^{1}\int_{0}^{1}\Pr\left[  Q_{t}\leq\frac{1}{x}\sqrt{\frac{2}%
{K}\ln\frac{1}{x}}\right]  dxdx\\
&  \leq4\int_{0}^{1}\int_{0}^{1}\min\left\{  1,\max\left\{  \frac{1}{x}%
\sqrt{\frac{2}{K}\ln\frac{1}{x}},\frac{1}{K}\right\}  g^{t/2}\right\}  dxdx\\
&  \leq4g^{t/2}\left(  \frac{1}{K}+\int_{0}^{1}\int_{0}^{1}\min\left\{
g^{-t/2},\frac{1}{x}\sqrt{\frac{2}{K}\ln\frac{1}{x}}\right\}  dxdx\right) \\
&  =4g^{t/2}\left(  \frac{1}{K}+\int_{x=0}^{1}\sqrt{\frac{2}{K}\ln\frac{1}{x}%
}\left(  x_{\min}\left(  x\right)  +\int_{x=x_{\min}\left(  x\right)  }%
^{1}\frac{1}{x}dx\right)  dx\right)
\end{align*}
Here the fifth line uses Lemma \ref{stillbig}, and%
\[
x_{\min}\left(  x\right)  =g^{t/2}\sqrt{\frac{2}{K}\ln\frac{1}{x}}.
\]
By straightforward integral approximations, the last expression is at most
$7g^{t/2+1}/\sqrt{K}$\ for sufficiently large $K$.
\end{proof}

For each child $w\in S\left(  v\right)  $, let%
\[
\eta_{t}\left(  w\right)  =\sum_{u\in\left\{  1,3,\ldots,t-1\right\}
}\left\vert \left\langle E_{B,u}\left(  w\right)  \right\rangle _{A}%
-E_{B,u}\left(  Y_{w}\right)  \right\vert
\]
measure the total error in Alice's estimates of $E_{B,u}\left(  Y_{w}\right)
$, summed over all time steps $u\leq t$. \ The following proposition shows
that to upper-bound the error in $\left\langle q_{t}\left(  w\right)
\right\rangle _{A}$, it suffices to upper-bound $\eta_{t}\left(  w\right)  $.
\ For this proposition to hold, we need the function $\Delta$\ to have bounded
derivative. \ That is why we chose triangular instead of uniform noise when
defining the protocol.

\begin{proposition}
\label{etaovere}%
\[
\left\vert \left\langle q_{t}\left(  w\right)  \right\rangle _{A}-q_{t}\left(
Y_{w}\right)  \right\vert \leq\frac{\eta_{t}\left(  w\right)  }{\epsilon}.
\]

\end{proposition}

\begin{proof}
From the definition of $\Delta$,%
\[
\left\vert \Delta\left(  m_{u+1},\left\langle E_{B,u}\left(  w\right)
\right\rangle _{A}\right)  -\Delta\left(  m_{u+1},E_{B,u}\left(  Y_{w}\right)
\right)  \right\vert \leq\frac{1}{\epsilon}\left\vert \left\langle
E_{B,u}\left(  w\right)  \right\rangle _{A}-E_{B,u}\left(  Y_{w}\right)
\right\vert .
\]
Furthermore, $\Delta\left(  m_{u+1},\left\langle E_{B,u}\left(  w\right)
\right\rangle _{A}\right)  $\ and $\Delta\left(  m_{u+1},E_{B,u}\left(
Y_{w}\right)  \right)  $\ are both bounded in $\left[  0,1\right]  $. \ It
follows that%
\begin{align*}
\left\vert \left\langle q_{t}\left(  w\right)  \right\rangle _{A}-q_{t}\left(
Y_{w}\right)  \right\vert  &  =\left\vert
%TCIMACRO{\dprod \limits_{u}}%
%BeginExpansion
{\displaystyle\prod\limits_{u}}
%EndExpansion
\Delta\left(  m_{u+1},\left\langle E_{B,u}\left(  w\right)  \right\rangle
_{A}\right)  -%
%TCIMACRO{\dprod \limits_{u}}%
%BeginExpansion
{\displaystyle\prod\limits_{u}}
%EndExpansion
\Delta\left(  m_{u+1},E_{B,u}\left(  Y_{w}\right)  \right)  \right\vert \\
&  \leq\sum_{u}\frac{1}{\epsilon}\left\vert \left\langle E_{B,u}\left(
w\right)  \right\rangle _{A}-E_{B,u}\left(  Y_{w}\right)  \right\vert
=\frac{\eta_{t}\left(  w\right)  }{\epsilon}%
\end{align*}
where $u$\ ranges over $\left\{  1,3,\ldots,t-1\right\}  $.
\end{proof}

Now let%
\begin{align*}
H  &  =\sum_{w\in S\left(  v\right)  }q_{t}\left(  Y_{w}\right)  ,\\
F  &  =\sum_{w\in S\left(  v\right)  }q_{t}\left(  Y_{w}\right)  f\left(
Y_{w}\right)  ,\\
\left\langle H\right\rangle _{A}  &  =\sum_{w\in S\left(  v\right)
}\left\langle q_{t}\left(  w\right)  \right\rangle _{A},\\
\left\langle F\right\rangle _{A}  &  =\sum_{w\in S\left(  v\right)
}\left\langle q_{t}\left(  w\right)  \right\rangle _{A}f\left(  Y_{w}\right)
,
\end{align*}
so that $E_{A,t}^{\ast}\left(  v\right)  =F/H$ and $\left\langle
E_{A,t}\left(  v\right)  \right\rangle _{A}=\left\langle F\right\rangle
_{A}/\left\langle H\right\rangle _{A}$. \ Using Lemma \ref{stillbig}, we can
upper-bound the probability that $H$ is too much smaller than its mean value.

\begin{corollary}
\label{stillbig2}For all $\gamma>0$,%
\[
\Pr_{y,M_{t},S\left(  v\right)  }\left[  H\leq\gamma K\right]  <3g^{t/2}%
\max\left\{  \gamma,\frac{4\ln K}{K}\right\}  .
\]

\end{corollary}

\begin{proof}
By the principle of deferred decisions, we can think of each $q_{t}\left(
Y_{w}\right)  $ as an independent sample of a $\left[  0,1\right]  $\ random
variable with mean $Q_{t}$. \ Then $H$\ is a sum of $K$\ such samples.
\ Setting $\Gamma=\max\left\{  2\gamma,8\left(  \ln K\right)  /K\right\}
$,\ by Lemma \ref{stillbig} we have%
\[
\Pr_{y,M_{t}}\left[  Q_{t}\leq\Gamma\right]  <2g^{t/2}\max\left\{
\gamma,\frac{4\ln K}{K}\right\}  .
\]
Furthermore, assuming $Q_{t}>\Gamma$, Theorem \ref{hoeffding} yields%
\[
\Pr_{S\left(  v\right)  }\left[  H\leq\gamma K\right]  \leq\exp\left(
-Q_{t}\left(  1-\frac{\gamma}{\Gamma}\right)  ^{2}\frac{K}{2}\right)  \leq
e^{-\Gamma K/8}\leq\frac{1}{K}.
\]
The corollary now follows by the union bound.
\end{proof}

The last piece of the puzzle is to upper-bound the difference between
$\left\langle E_{A,t}\left(  v\right)  \right\rangle _{A}$ and $E_{A,t}^{\ast
}\left(  v\right)  $, using techniques similar to those of Lemma
\ref{integrals}. \ Let $\eta=\operatorname*{EX}_{w\in S\left(  v\right)
}\left[  \eta_{t}\left(  w\right)  \right]  $ and $\widehat{\eta
}=\operatorname*{EX}_{y,M_{t},\mathcal{T}_{A}}\left[  \eta\right]  $.

\begin{lemma}
\label{finally}Assuming $\eta\geq1/K$ for all $y,M_{t},\mathcal{T}_{A}$,%
\[
\operatorname*{EX}_{y,M_{t},\mathcal{T}_{A}}\left[  \left\vert \left\langle
E_{A,t}\left(  v\right)  \right\rangle _{A}-E_{A,t}^{\ast}\left(  v\right)
\right\vert \right]  \leq18g^{t/2+1}\widehat{\eta}.
\]

\end{lemma}

\begin{proof}
Using the fact that $\left\langle F\right\rangle _{A}\leq\left\langle
H\right\rangle _{A}$,%
\[
\left\vert \left\langle E_{A,t}\left(  v\right)  \right\rangle _{A}%
-E_{A,t}^{\ast}\left(  v\right)  \right\vert =\left\vert \frac{\left\langle
F\right\rangle _{A}}{\left\langle H\right\rangle _{A}}-\frac{F}{H}\right\vert
\leq\frac{\left\vert \left\langle H\right\rangle _{A}-H\right\vert }{H}%
+\frac{\left\vert \left\langle F\right\rangle _{A}-F\right\vert }{H}%
\]
by the same trick as in Corollary \ref{chernoffcor}. \ Furthermore, it follows
from Proposition \ref{etaovere}\ together with the triangle inequality that
$\left\vert \left\langle H\right\rangle _{A}-H\right\vert \leq\eta K/\epsilon
$\ and $\left\vert \left\langle F\right\rangle _{A}-F\right\vert \leq\eta
K/\epsilon$. \ So we can upper-bound $\left\vert \left\langle E_{A,t}\left(
v\right)  \right\rangle _{A}-E_{A,t}^{\ast}\left(  v\right)  \right\vert $ by
$2\eta K/\left(  \epsilon H\right)  $, as well as (of course) by $1$. Fix
$\eta$; then%
\begin{align*}
\operatorname*{EX}_{H}\left[  \min\left\{  1,\frac{2\eta K}{\epsilon
H}\right\}  \right]   &  =\int_{0}^{1}\Pr_{H}\left[  \frac{2\eta K}{\epsilon
H}\geq x\right]  dx\\
&  \leq\int_{0}^{1}\min\left\{  1,3\max\left\{  \frac{2\eta}{\epsilon x}%
,\frac{4\ln K}{K}\right\}  g^{t/2}\right\}  dx\\
&  \leq3g^{t/2}\left(  \frac{4\ln K}{K}+\int_{0}^{1}\min\left\{  \frac
{1}{3g^{t/2}},\frac{2\eta}{\epsilon x}\right\}  dx\right) \\
&  =3g^{t/2}\left(  \frac{4\ln K}{K}+\frac{x_{\min}}{3g^{t/2}}+\int_{x_{\min}%
}^{1}\frac{2\eta}{\epsilon x}dx\right)
\end{align*}
where the second line uses Corollary \ref{stillbig2} and $x_{\min}=\left(
6\eta/\epsilon\right)  g^{t/2}$. \ This in turn is at most%
\[
3g^{t/2}\left(  \frac{4\ln K}{K}+\frac{2\eta}{\epsilon}\ln\frac{1}{\eta
}\right)  .
\]
Assuming $\eta\geq1/K$\ always, the expectation of the above quantity over
$\eta$\ is at most $18g^{t/2+1}\widehat{\eta}$.
\end{proof}

We are finally ready to put everything together, and show that the referee can
distinguish the real and ideal conversations with bias at most $\zeta$.

\begin{theorem}
\label{efficient}By setting $b=\left\lceil \log_{2}R/\left(  \zeta
\epsilon\right)  \right\rceil +2$\ and $K=O\left(  \left(  11/\epsilon\right)
^{R^{2}}/\zeta^{2}\right)  $, it is possible to achieve%
\[
\left\vert \Pr_{\omega\in\mathcal{D},M\in\mathcal{W}\left(  \omega\right)
}\left[  \Phi\left(  \omega,M_{R}\right)  =1\right]  -\Pr_{\omega
\in\mathcal{D},M\in\mathcal{B}\left(  \omega\right)  }\left[  \Phi\left(
\omega,M_{R}\right)  =1\right]  \right\vert \leq\zeta
\]
for all Boolean functions $\Phi$.
\end{theorem}

\begin{proof}
Combining Lemmas \ref{integrals} and \ref{finally},%
\[
\operatorname*{EX}_{y,M_{t},\mathcal{T}_{A}}\left[  \left\vert \left\langle
E_{A,t}\left(  v\right)  \right\rangle _{A}-E_{A,t}\left(  X_{v}\right)
\right\vert \right]  \leq g^{t/2+1}\left(  \frac{7}{\sqrt{K}}+18\widehat{\eta
}\right)  .
\]
Let $\mathcal{L}_{j}$\ be the set of nodes at the $j^{th}$\ level of Alice's
tree $\mathcal{T}_{A}$. \ Then if $j$ is even, let
\begin{align*}
\lambda_{j}  &  =\operatorname*{EX}_{v\in\mathcal{L}_{j}}\left[  \sum
_{t\in\left\{  j,j+2,\ldots,R\right\}  }\operatorname*{EX}_{y,M_{t}%
,\mathcal{T}_{A}}\left[  \left\vert \left\langle E_{A,t}\left(  v\right)
\right\rangle _{A}-E_{A,t}\left(  X_{v}\right)  \right\vert \right]  \right]
,\\
\lambda_{j+1}  &  =\operatorname*{EX}_{w\in\mathcal{L}_{j+1}}\left[
\sum_{t\in\left\{  j+1,j+3,\ldots,R-1\right\}  }\operatorname*{EX}%
_{y,M_{t},\mathcal{T}_{A}}\left[  \left\vert \left\langle E_{B,t}\left(
w\right)  \right\rangle _{A}-E_{B,t}\left(  Y_{w}\right)  \right\vert \right]
\right]  .
\end{align*}
By linearity of expectation,%
\[
\lambda_{j}\leq\left(  \frac{R}{2}+1\right)  g^{R/2+1}\left(  \frac{7}%
{\sqrt{K}}+18\lambda_{j+1}\right)  .
\]
Solving this recurrence relation, we find that at the root node,%
\[
\lambda_{0}\leq\left(  9R+18\right)  ^{R}g^{R^{2}/2+R}\frac{7}{\sqrt{K}},
\]
and similarly for the root of Bob's tree $\mathcal{T}_{B}$. \ So in
particular, $\operatorname*{EX}_{\omega,M_{R},\mathcal{T}_{i}}\left[
\partial_{t}\right]  \leq\lambda_{0}+2^{-b+1}$\ for all $i,t$, where%
\[
\partial_{t}=\left\vert \operatorname*{round}\left(  \left\langle
E_{i,t}\left(  \operatorname*{root}\nolimits_{i}\right)  \right\rangle
_{i}\right)  -\operatorname*{round}\left(  E_{i,t}\left(  \omega\right)
\right)  \right\vert .
\]
Now observe that, if we let $\mathcal{W}_{t+1}$\ be the distribution over
message $m_{t+1}$\ in the wannabe case, and let $\mathcal{B}_{t+1}$\ be the
distribution in the unbounded Bayesian case, then%
\[
\left\Vert \mathcal{W}_{t+1}-\mathcal{B}_{t+1}\right\Vert _{1}=\frac{1}{2}%
\sum_{r=1}^{\partial_{t}/2^{-b}}\frac{2\left(  L-r+1\right)  }{L^{2}}\leq
\frac{\partial_{t}/2^{-b}}{L}=\frac{\partial_{t}}{\epsilon}%
\]
where $\left\Vert ~\right\Vert _{1}$\ denotes variation distance. \ So the
referee can distinguish the whole conversations with bias at most%
\[
\frac{1}{\epsilon}\operatorname*{EX}\left[  \partial_{0}+\cdots+\partial
_{R-1}\right]  \leq\frac{1}{\epsilon}\left(  \lambda_{0}+2^{-b+1}\right)  R
\]
since variation distance satisfies the triangle inequality. \ Therefore, we
can achieve the goal of simulation by taking $\lambda_{0}\leq\zeta
\epsilon/R-2^{-b+1}\leq\zeta\epsilon/2R$, or equivalently%
\[
K=\frac{196R^{2}}{\zeta^{2}\epsilon^{2}}\left(  9R+18\right)  ^{2R}\left(
\frac{4e}{\epsilon}\right)  ^{R^{2}+2R}\left(  \ln\left(  \frac{196R^{2}%
}{\zeta^{2}\epsilon^{2}}\left(  9R+18\right)  ^{2R}\left(  \frac{4e}{\epsilon
}\right)  ^{R^{2}+2R}\right)  \right)  ^{2R}=O\left(  \frac{1}{\zeta^{2}%
}\left(  \frac{11}{\epsilon}\right)  ^{R^{2}}\right)  .
\]

\end{proof}

\section{Discussion\label{OPEN}}

\begin{quote}
\textquotedblleft We publish this observation with some diffidence, since once
one has the appropriate framework, it is mathematically trivial.
\ Intuitively, though, it is not quite obvious\ldots\textquotedblright%
\ ---Aumann \cite{aumann}, on his original agreement result
\end{quote}

This paper has studied agreement protocols from the quantitative perspective
of theoretical computer science. \ If nothing else, we hope to have shown that
adopting that perspective leads to rich mathematical questions. \ Here are a
few of the more interesting open problems raised by our results:

\begin{itemize}
\item How tight is our $O\left(  1/\left(  \delta\varepsilon^{2}\right)
\right)  $ upper bound? \ Can we improve Theorem \ref{dspthm}\ to show that
the discretized standard protocol uses only $O\left(  1/\varepsilon
^{2}\right)  $\ messages, independently of $\delta$? \ More importantly, is
there a scenario where Alice and Bob must exchange $\Omega\left(
1/\varepsilon\right)  $\ or $\Omega\left(  1/\varepsilon^{2}\right)  $\ bits
to $\left(  \varepsilon,1/2\right)  $-agree, regardless of what protocol they
use? \ Recall that the best lower bound we currently know is $\Omega\left(
\log1/\varepsilon\right)  $, from Proposition \ref{logprop}.

\item Can Alice and Bob $\left(  \varepsilon,\delta\right)  $-agree after a
small number of steps, even if the \textquotedblleft true\textquotedblright%
\ distribution over $\omega$ differs from their shared prior distribution
$\mathcal{D}$? \ Or is there a scenario where regardless of what protocol they
use, there exists a state $\omega$ for which they must exchange $\Omega\left(
n\right)  $\ bits to agree within $\varepsilon$ on $\omega$? \ (It is easy to
construct a scenario where the discretized standard protocol needs
$\Omega\left(  n\right)  $\ bits for some $\omega$.)

\item Can the simulation procedure of Section \ref{SIMSMOOTH}\ be made
practical? \ That is, can we reduce the number of subroutine calls to (say)
$c^{1/\left(  \delta\varepsilon^{2}\right)  }$, or even to a polynomial in
$1/\delta$\ and $1/\varepsilon$? \ Alternatively, can we prove a lower bound
showing that such reductions are impossible?

\item Can we obtain a better simulation procedure if $\mathcal{D}$\ is
represented in a compact form, for example a graphical model?
\end{itemize}

Stepping back, have the results of this paper taught us anything about the
origins of disagreement? \ As mentioned in Section \ref{INTRO}, it is easy to
list plausible reasons why people might disagree, Aumann's theorem
notwithstanding: indifference to truth, misconstrual, vagueness, dishonesty,
self-deceit, mistrust, stupidity, systematic cognitive biases, no priors,
different priors, different indexicality assumptions, diagonalization (as
discussed in Section \ref{CC}), communication cost, and computation cost,
among others. \ But which of these reasons, if any, are fundamental? \ In
other words, were we forced to identify a single point at which the
assumptions of Aumann's theorem diverge from reality, what would it be?

Before we undertook the research described in this paper, we would have said
\textit{either} that

\begin{enumerate}
\item[(1)] imposing reasonable communication and computation bounds is likely
to change everything, or

\item[(2)] at least one party to any persistent disagreement must be
dishonest, irrational, or indifferent to truth.\footnote{Here
\textquotedblleft indifference to truth\textquotedblright\ means choosing
opinions according to their novelty, social acceptability, value in attracting
sexual partners, etc. rather than evidence.}
\end{enumerate}

Today, however, we would make an argument less technical than (1) and less
misanthropic than (2): that even in idealized models,\textit{ we should not
treat agents as initially-identical Bayesian \textquotedblleft
containers\textquotedblright\ that later get filled with different
experiences.} \ In particular, the Common Prior Assumption (CPA) is
fundamentally misguided.

Presumably no one would claim that the CPA is empirically true for human
beings. \ It seems obvious that, when five-year-olds go to Sunday school, they
are not updating a shared prior over possible religions conditioned on what
their teacher tells them. \ Rather, their priors are being \textquotedblleft
initialized\textquotedblright\ to some extent. \ Furthermore, the existence of
a common prior would be astonishing from the perspectives of physics,
evolutionary biology, and neuroscience, since nothing in those fields predicts
or requires one. \ However, as Aumann \cite{aumann2}\ rightly emphasizes, the
question is not whether the CPA is \textquotedblleft true\textquotedblright%
\ but whether it is a useful idealization. \ What we suggest is that, when
trying to understand the origins of disagreement,\ the CPA is \textit{not} a
useful idealization. \ There are two main reasons for this.

First, the CPA presents difficulties with transtemporal identity. \ Are you
really the \textquotedblleft same\textquotedblright\ person as you were when
you were two months old? \ If not, then why must your posterior be obtained by
updating the two-month-old's prior? \ The difficulties become even more severe
if we adopt the many-worlds view of quantum mechanics. \ For then there are
millions of basis states containing beings very much like you. \ Suppose we
fix which one of those beings is \textquotedblleft really\textquotedblright%
\ you at time $t$; then which one is you at times $t-1$\ or $t+1$?\ \ Quantum
mechanics does not fix an answer; more than that, it does not even fix the
probabilities of \textit{possible} answers.\footnote{What quantum mechanics
does fix are the probabilities of possible outcomes of a measurement. \ But
those probabilities will only be meaningful to you if you are not part of the
system being measured.} \ That is why Bohmian mechanics and its many variants
can all be compatible with quantum mechanics, despite having different
equations of motion.

Second, the CPA begs the question of what determines the common prior. \ Some
might argue that human beings' shared genetic heritage causes them (or rather,
should cause them) to share a prior. \ But if your prior is to fix your
initial opinions about \textit{everything}, then it must assign a probability
to your future experiences being consistent with those of (say) a five-legged
extraterrestrial. \ Presumably that probability decreases dramatically once
you condition on the indexical fact of your humanity. \ But it ought to start
nonzero and stay nonzero, for instance because of quantum fluctuations. \ This
raises a question: why shouldn't your prior \textit{equal} the
extraterrestrial's? \ After all, the extraterrestrial has to assign a
probability to \textit{its} future experiences being consistent with
yours---and at a hypothetical time before either of you knows who
\textquotedblleft you\textquotedblright\ will become, why should the two of
you reason differently? \ We can similarly imagine beings governed by
different laws of physics; and these, too, should share our prior. \ It
follows that \textquotedblleft the\textquotedblright\ common prior, if it
exists, is not determined by anything in our genetic makeup or even the
physical world.

This leaves the possibility that mathematics or logic could determine the
common prior. \ Along these lines, Schmidhuber\ \cite{schmidhuber}\ has
advocated a prior\ in which the probability of any sequence of experiences $x$
is proportional to $2^{-K\left(  x\right)  }$, where $K$ is the Kolmogorov
complexity of $x$---that is, the length of the shortest computer program that
outputs $x$. \ This idea has several problems, though. \ First, our actual
experiences seem to have gratuitously high Kolmogorov complexity. \ Believers
in the Kolmogorov prior are forced to say, without evidence, that this is an
illusion. \ Second, why should we use Kolmogorov complexity, rather than (say)
time-bounded Kolmogorov complexity, or perhaps the length of the shortest
program that outputs $x$ given an oracle for the halting problem? \ Third,
whenever we wish to compare the probabilities of a few \textquotedblleft
equally complex\textquotedblright\ events, the probabilities will depend less
on the events themselves than on our choice of programming language, so we
face another arbitrary choice.

So it seems that a common prior would be independent of the physical world and
even of mathematics, yet would somehow be readily available to and
unquestioningly accepted by every rational agent. \ Agents equipped with this
prior would live a `preprogrammed' existence, meaning that they would never
change, only conditionalize. \ We have argued that this picture of the world
presents serious intrinsic problems, even setting aside its naked
implausibility. \ So perhaps the common prior should be jettisoned with the ether.

But is there any principled basis for prior differences, then?\ \ Consider
Shakespeare's Julius Caesar, debating whether to venture outside on the Ides
of March. \ From his dismissals of omens, we know that Caesar bases his final
decision on a \textit{belief} that he will not be in particular danger, rather
than just a \textit{preference} for risky actions. \ Yet the process of
reaching the belief\ seems to have nothing to do with conditioning on
evidence---or rather, it starts after the conditioning is already done. \ Our
proposal is to view the process as that of Caesar \textit{choosing his prior,
and thereby choosing what sort of person he is}. \ In other words, Caesar
assigns a low prior probability to his getting killed, for the sole reason
that had he assigned a high one, he would no longer be Caesar but someone
else.\footnote{[D]anger knows full well
\par
That Caesar is more dangerous than he:
\par
We are two lions litter'd in one day,
\par
And I the elder and more terrible...
\par
---\textit{Julius Caesar}, Act 2, Scene 2} \ On this view, not only can Alice
and Bob have different priors because they are different people, but the fact
that they have different priors is a large part of what \textit{makes} them
different people, rather than the same person filling two pairs of shoes.

In saying this, we are not taking the relativist stance that any prior is
\textquotedblleft rational\textquotedblright\ for the sort of person who would
hold that prior. \ If no priors are objectively more rational than others,
then the word \textquotedblleft rational\textquotedblright\ is meaningless,
since there exists a prior to justify essentially any\ belief. \ But the
question remains: is the number of rational priors exactly one? \ We have
already seen an argument of Hanson \cite{hanson2} that it should be, based on
the concept of a \textquotedblleft pre-prior\textquotedblright\ (that is, a
prior over all possible priors). \ Why should Alice give her own prior any
more weight than Bob's? \ Our response is simply to point out that there is a
tremendous gap between empathizing with someone else's perspective and
adopting it, or between calculating what your expectation would be under
someone else's prior and willing that expectation to be yours. \ No matter how
long she talks to Bob, in the end Alice must confront the irreducible fact of
her individuality. \ As Clarence Darrow famously put it, \textquotedblleft I
don't like spinach, and I'm glad I don't, because if I liked it I'd eat it,
and I just hate it.\textquotedblright

\section{Acknowledgments}

I thank Robin Hanson for correspondence during the early stages of this work,
and for his wonderfully disorienting papers which introduced me to the topic.
\ I also thank Ran Raz, Umesh Vazirani, and Ronald de Wolf for helpful conversations.

\end{document}